\documentclass[12pt,preprint,useAMS,usenatbib,usegraphicx,english]{aastex}

\usepackage[T1]{fontenc}
\usepackage[utf8]{inputenc}
\usepackage{babel}
\shortauthors{CASTRO ET AL.}
\shorttitle{HIGH PROPER MOTION L DWARFS}
\begin{document}

\title{DISCOVERY OF FOUR HIGH PROPER MOTION L DWARFS, INCLUDING A 10 PC L DWARF AT THE L/T TRANSITION\thanks{Observations 
reported here were obtained at the MMT Observatory, a joint facility of the University of Arizona and the Smithsonian 
Institution. MMT telescope time was granted by NOAO, through the Telescope System Instrumentation Program (TSIP). TSIP 
is funded by NSF.}
$^{,}$\thanks{Based on observations obtained with the Apache Point Observatory 3.5-meter telescope, which is owned and operated 
by the Astrophysical Research Consortium.}}

\author{Philip J. Castro\thanks{Visiting Astronomer at the Infrared Telescope Facility,
which is operated by the University of Hawaii under Cooperative Agreement no. NNX-08AE38A with the National
Aeronautics and Space Administration, Science Mission Directorate, Planetary Astronomy Program.} ~and John E. Gizis\footnotemark[3]}
\affil{Department of Physics and Astronomy, University of Delaware, Newark, DE 19716, USA; pcastro@udel.edu, gizis@udel.edu}
\author{Hugh C. Harris}
\affil{US Naval Observatory, Flagstaff Station, 10391 West Naval Observatory Road, Flagstaff, AZ 86001, USA}
\author{Gregory N. Mace}
\affil{Department of Physics and Astronomy, UCLA, Los Angeles, CA 90095-1547, USA}
\author{J. Davy Kirkpatrick}
\affil{Infrared Processing and Analysis Center, MS 100-22, California Institute of Technology, Pasadena, CA 91125, USA}
\author{Ian S. McLean}
\affil{Department of Physics and Astronomy, UCLA, Los Angeles, CA 90095-1547, USA}
\author{Petchara Pattarakijwanich}
\affil{Department of Astrophysical Sciences, Princeton University, Ivy Lane, Princeton, NJ 08544, USA}
\author{Michael F. Skrutskie}
\affil{Department of Astronomy, University of Virginia, Charlottesville, VA 22904, USA}

\begin{abstract}
We discover four high proper motion L dwarfs by comparing the Wide-field Infrared Survey Explorer (WISE) to the 
Two Micron All Sky Survey (2MASS). WISE J140533.32+835030.5 is an L dwarf at the L/T transition with a 
proper motion of $0.85\pm0\farcs02$ yr$^{-1}$, previously overlooked due to its proximity to a bright star (V$\approx$12 mag). 
From optical spectroscopy we find a spectral type of L8, and from moderate-resolution $J$ band spectroscopy we find a 
near-infrared spectral type of L9. We find WISE J140533.32+835030.5 to have a distance of $9.7\pm1.7$ pc, bringing the number of L dwarfs 
at the L/T transition within 10 pc from six to seven. WISE J040137.21+284951.7, WISE J040418.01+412735.6, 
and WISE J062442.37+662625.6 are all early L dwarfs within 25 pc, and were classified using optical and low-resolution 
near-infrared spectra. WISE J040418.01+412735.6 is an L2 pec (red) dwarf, a member of the class of unusually red L dwarfs. We use follow-up 
optical and low-resolution near-infrared spectroscopy to classify a previously discovered \citep{CastroGizis2012} fifth 
object WISEP J060738.65+242953.4 as an (L8 Opt/L9 NIR), confirming it as an L dwarf at the L/T transition within 10 pc. 
WISEP J060738.65+242953.4 shows tentative CH$_{4}$ in the $H$ band, possibly the result of unresolved binarity with an early T dwarf, 
a scenario not supported by binary spectral template fitting. If WISEP J060738.65+242953.4 is a single object, it represents the earliest onset of CH$_{4}$ 
in the $H$ band of an L/T transition dwarf in the SpeX Library. As very late L dwarfs within 10 pc, WISE J140533.32+835030.5 and WISEP J060738.65+242953.4 
will play a vital role in resolving outstanding issues at the L/T transition. 
\end{abstract}


\section{INTRODUCTION}
The Wide-field Infrared Survey Explorer (WISE) all-sky data release occurred on March 14, 2012.
The survey covers the entire sky in four bands centered at wavelengths 3.4$\mu$m ($W1$), 4.6$\mu$m ($W2$), 12$\mu$m ($W3$), 
and 22$\mu$m ($W4$), and achieves 5$\sigma$ detections for point sources. One of the primary science goals of WISE is 
to search for cool brown dwarfs, T dwarfs to Y dwarfs, with the $W1-W2$ color playing an essential role due to a 
lack of methane absorption at the 4.6$\mu$m band relative to the 3.4$\mu$m band \citep{Kirkpatrick2005,Wrightetal2010}. 
WISE has yielded numerous T dwarf discoveries, from the first few T dwarfs by \citet{Burgasseretal2011b} and \citet{Mainzeretal2011}, 
to the discovery of the first $\approx$ 100 by \citet{Kirkpatricketal2011} and 87 new T dwarfs by \citet{Maceetal2013}, 
culminating with the extension of the spectral class from T to Y with the first detection of Y dwarfs \citep{Cushingetal2011,Kirkpatricketal2012}.
The Two Micron All Sky Survey (2MASS) is a near-infrared survey performed from 1997 to 2001 covering virtually the entire sky at wavelengths 
1.25$\mu$m ($J$), 1.65$\mu$m ($H$), and 2.16$\mu$m ($K_{\rm s}$) \citep{Skrutskieetal2006}. A consequence of two all-sky surveys with wavelengths 
in the near-infrared and mid-infrared, with a difference in epochs of $\sim$ 10 yr, is that it creates an ideal setup to search for 
ultracool dwarfs with large proper motion.
Multi-epoch searches using WISE have proven successful at discovering high proper 
motion ultracool dwarfs \citep{Aberasturietal2011,Liuetal2011,Loutreletal2011,Gizisetal2011a,Gizisetal2011b,Scholzetal2011,CastroGizis2012,Gizisetal2012,Luhmanetal2012,Luhman2013}.

Discoveries with WISE have yielded significant increases in the number of nearby ($\leq$ 10 pc) very late L ($\geq$L7) dwarfs,
a demographic that is rare \citep{CastroGizis2012}. Prior to the WISE preliminary data release this population consisted 
of only two; the (L8 Opt/L9 NIR) (L$_{\rm o}$8/L$_{\rm n}$9) \citep{Kirkpatricketal2008,Burgasseretal2006a} dwarf 
DENIS-P J0255-4700 \citep{Martinetal1999} at $4.97\pm0.10$ pc \citep{Costaetal2006}, and the 
L$_{\rm o}$8 \citep{Kirkpatricketal2008} dwarf 2MASS J02572581-3105523 \citep{Kirkpatricketal2008} 
at 9.7$\pm$1.3 pc \citep{Looperetal2008b}. From WISE discoveries;
the L$_{\rm o}$8$\pm1$/L$_{\rm n}$7.5 \citep{Luhman2013,Burgasseretal2013} dwarf WISE J104915.57-531906.1A \citep{Luhman2013} 
at 2.0$\pm$0.15 pc \citep{Luhman2013},
the L$_{\rm o}$8 dwarf WISEP J060738.65+242953.4 
at 7.8$^{+1.4}_{-1.2}$ pc \citep{CastroGizis2012}, the L$_{\rm n}$9 pec (red) dwarf WISEPA J164715.59+563208.2 
at 8.6$^{+2.9}_{-1.7}$ pc \citep{Kirkpatricketal2011}, and the L$_{\rm n}$7.5 dwarf WISEP J180026.60+013453.1 
at 8.8$\pm$1.0 pc \citep{Gizisetal2011a}. WISE discoveries have tripled the number of very late L dwarfs 
within 10 pc. 

The L/T transition occurs over a small temperature span of $\sim$200-300 K 
at T$_{\rm eff}\approx$1500 K \citep{Kirkpatrick2005} and over a relatively short period of 
time ($\sim100$ Myr for a 0.03 $M_{\odot}$ brown dwarf) \citep{Burgasser2007}. The L/T transition is believed to be caused by the 
depletion of condensate clouds, where the driving mechanism for the depletion is inadequately explained by current 
cloud models \citep{Burgasseretal2011a}. The bluer $J-K$ and the brightening of the $J$ band at the L/T transition 
can be explained by decreasing cloudiness. A mechanism suggested for the L/T dwarf spectral type transition is the 
appearance of relatively cloud free regions across the disk of the dwarfs as they cool \citep{Marleyetal2010}. The 
complex dynamic behavior of condensate clouds of low temperature atmospheres at the L/T transition is one of the 
leading problems in brown dwarf astrophysics today \citep{Burgasseretal2011a}.

We present the discovery of four L dwarfs within 25 pc, as part of an ongoing effort to discover high proper 
motion objects between 2MASS and WISE \citep{Gizisetal2011a,Gizisetal2011b,Gizisetal2012,CastroGizis2012}. In Section 2 
we present our analysis, discussing the discovery of each object in turn, their proper motion, and determine a transformation 
between $I_{\rm C}$ and $i$. In Section 3 we present our observations. In Section 4 we present the spectral 
analysis of the four newly discovered L dwarfs. In Section 5 we present follow-up spectroscopy confirming 
WISEP J060738.65+242953.4 as an L dwarf at the L/T transition. Lastly, in Section 6 we present our conclusions and discuss 
future work.

\section{ANALYSIS}
Our analysis first discusses the optical imaging of WISE J140533.32+835030.5 (W1405+8350), from which we determine
position and photometry. We then discuss the discovery of the L dwarfs, the general search strategy and the discovery 
of each L dwarf in detail, followed by their proper motion. Lastly, we determine a transformation 
between $I_{\rm C}$ and $i$ for L dwarfs.

\subsection{Optical Imaging}
W1405+8350 was observed on UT Date 23 March 2012 with the 1.55 m Strand Astrometric Reflector at the Flagstaff Station of
the US Naval Observatory (USNO) using a Tek2K 2048x2048 CCD Camera. We obtained $I_{\rm C}$ and $z$ band images of W1405+8350.
The data were reduced using standard techniques and calibrated to the known Sloan Digital Sky Survey
magnitude ($z$ = 15.28 AB mag, \citealt{Aiharaetal2011}) of SDSS J152702.74+434517.2, observed immediately beforehand. 
The observed magnitudes of W1405+8350 are $z$=17.50$\pm$0.04 and $I_{\rm C}$=18.93$\pm$0.06.

\subsection{Discovery}
We used similar criteria to search for high proper motion objects as \citet{Gizisetal2011b}, but extended the search 
to red colors. 
We searched for WISE sources that had detections at $W1$ (3.4$\mu$m), $W2$ (4.6$\mu$m), and $W3$ (12$\mu$m), 
and no 2MASS counterpart within $3^{\prime \prime}$.
Sources in the WISE catalog are already matched to 2MASS sources 
within $3^{\prime \prime}$. By selecting sources without a 2MASS counterpart, with a difference in all-sky surveys 
of about a decade, this constrained the search to sources with a proper motion of $>0\farcs3$ yr$^{-1}$.
WISE sources meeting our criteria were examined visually using 2MASS and WISE 
finder charts\footnote{Finder charts at IRSA can be found at http://irsa.ipac.caltech.edu/} in order to 
look for apparent motion of a source between the two surveys.
The discoveries presented here are only a subset of the entire survey, the full catalog 
of high proper motion objects will be reported in a future paper, Gizis \& Castro, in prep.
WISE J040137.21+284951.7, WISE J040418.01+412735.6, and WISE J062442.37+662625.6 were discovered from the WISE preliminary
data release\footnote{http://wise2.ipac.caltech.edu/docs/release/prelim/expsup/} and are reanalyzed using the WISE all-sky
data release\footnote{http://wise2.ipac.caltech.edu/docs/release/allsky/expsup/}, and WISE J140533.32+835030.5 was discovered 
from the WISE all-sky data release. 

WISE J140533.32+835030.5 (W1405+8350) was found to have a separation of $\approx8.5^{\prime \prime}$ from a 2MASS 
source to the southeast, 2MASS J14053729+8350248. 2MASS J14053729+8350248 was previously overlooked due to its 
proximity ($\approx11^{\prime \prime}$) to a bright background star, 2MASS J14053168+8350188 ($J$=10.71 and 
$V$=11.75, \citealt{Laskeretal2008}), resulting in a photometric confusion flag for the $J$ band (`cc\_flg' of `c00').
The object would otherwise have been identified in previous searches of 2MASS for ultracool dwarfs \citep{Kirkpatricketal2000,Cruzetal2003}.
In our USNO $z$ band imaging data, J140532.57+835031.7 was detected $\approx2^{\prime \prime}$ to the northwest of W1405+8350, 
along the line of motion between the 2MASS and WISE sources, confirming W1405+8350 as a high proper motion object.
The WISE source shows colors that are red, $W1-W2=0.58\pm0.04$, consistent with that of a late L dwarf/early 
T dwarf \citep{Kirkpatricketal2011}; the 2MASS source has red colors, $J-K_{\rm s}=1.78\pm0.06$,
that are consistent with an L dwarf \citep{Kirkpatricketal2000}; and the USNO source has red colors, $I_{\rm C}-z=1.43\pm0.07$, 
consistent with a late L dwarf \citep{Dahnetal2002}. We positively identify the 2MASS and USNO sources as W1405+8350 at 
their respective epochs. With a high proper motion indicating a nearby object and red colors in 2MASS, WISE, and USNO 
indicating a late spectral type, we claim the detection of a nearby ultracool dwarf. A finder chart for W1405+8350 showing 
a clear linear sequence of positions at the epoch of 2MASS, WISE, and USNO is shown in Figure 1.

\begin{figure}
\caption{
Finder chart showing the proper motion of W1405+8350 from the 2MASS $K_{\rm s}$ band image (top left) to the WISE $W1$ 
image (top right) to the USNO $z$ band image (bottom left). Each image is 4$^{\prime}$ x 4$^{\prime}$, the circle shows the 
position of W1405+8350. The three 40$^{\arcsec}$ x 40$^{\arcsec}$ tiles in the bottom right show zoomed in images with the 
same configuration as the larger images. The three circles in each image show, from bottom left to top right, the position 
of W1405+8350 at the 2MASS, WISE, and USNO positions, respectively. North is up and east is to the left.
}
\begin{center}
\includegraphics[width=6.5in]{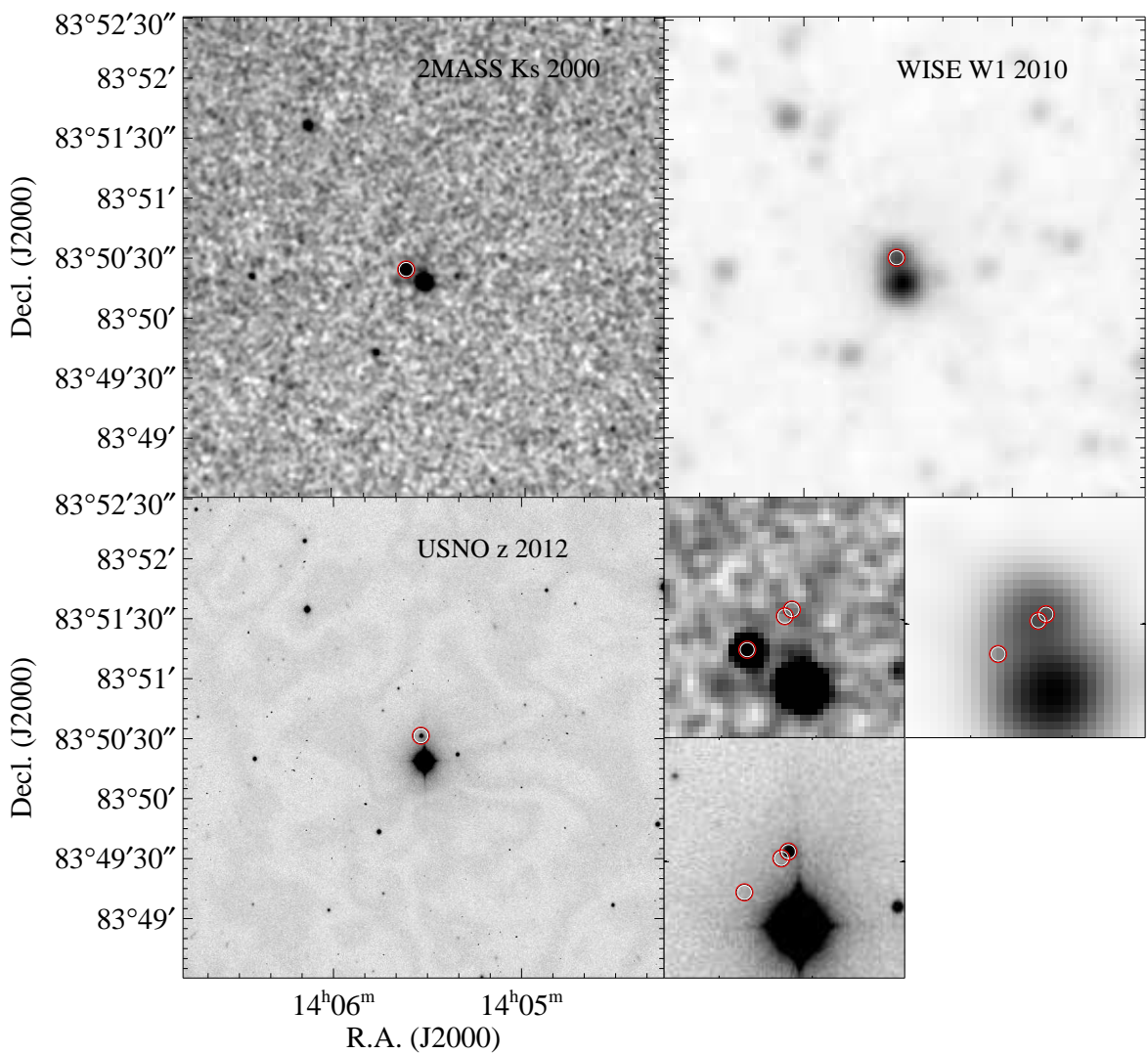}
\end{center}
\end{figure}

WISE J040137.21+284951.7 (W0401+2849) was found to have a 
separation of $\approx6\arcsec$ from a 2MASS source to the east, 2MASS J04013766+2849529. 
W0401+2849 is detected on several of the Palomar Observatory Sky Survey (POSS) plates, POSS2 red and POSS2 IR, with at
best a mediocre detection on POSS1 red, confirming the apparent motion between 2MASS and WISE. See Figure 2 for the finder chart.
The WISE source shows colors that 
are red, $W1-W2=0.27\pm0.03$, consistent with that of a late M dwarf/early L dwarf \citep{Kirkpatricketal2011}; the 2MASS 
source has red colors, $J-K_{\rm s}=1.59\pm0.03$, that are consistent with an early L dwarf \citep{Kirkpatricketal2000}.
This source was just outside of the $J$ band magnitude cut ($<20$ pc) in a search for ultracool dwarfs in the 2MASS 
Second Incremental Release by \citet{Cruzetal2003}.

WISE J040418.01+412735.6 (W0404+4127) was found to have a separation of $\approx4\arcsec$ 
from a 2MASS source to the north, 2MASS J04041807+4127398. 
W0404+4127 is detected on several of the POSS plates, POSS2 red and POSS2 IR, confirming the apparent motion between 2MASS and WISE.
It is however too faint to be detected on the POSS1 red plate. See Figure 2 for the finder chart.
The WISE source shows colors that are red, $W1-W2=0.30\pm0.03$,
consistent with that of a late M dwarf/early L dwarf \citep{Kirkpatricketal2011}; the 2MASS source has red colors, 
$J-K_{\rm s}=1.73\pm0.04$, that are consistent with an L dwarf \citep{Kirkpatricketal2000}.
This source was likely overlooked because it lies at a low galactic latitude, $b=-8^\circ$, where previous searches
often avoided the galactic plane \citep{Gizisetal2000,Cruzetal2003}.

WISE J062442.37+662625.6 (W0624+6626) 
was found to have a separation of $\approx6.5\arcsec$ from a 2MASS source to the northwest, 2MASS J06244172+6626309. 
2MASS J06244172+6626309 is reported as a source having poor photometry with 2MASS `PH\_QUAL'\footnote{The `PH\_QUAL' flag is a measure of the 
photometric quality in each band, with flags A, B, C, D, E, F, U, and X. A to C represent detections with a decreasing signal 
to noise. For more details refer to the 2MASS All-Sky Data Release Explanatory 
Supplement, http://www.ipac.caltech.edu/2mass/releases/allsky/doc/explsup.html} flags of `EEA' for the $JHK_{\rm s}$ bands, 
respectively.
W0624+6626 was originally reported as an extended source in the WISE preliminary data release, 
WISEP J062442.27+662626.1, with the all-sky release resolving it into two distinct sources, W0624+6626 and 
WISE J062441.53+662629.2. WISE J062441.53+662629.2 lies on the position of the 2MASS source to the northwest, 
2MASS J06244172+6626309, with $W1-W2\approx0$, it is likely earlier than M0 \citep{Kirkpatricketal2011}. WISE J062441.53+662629.2 
appears to be a background star that W0624+6626 was in close proximity to during the 2MASS epoch, resulting in poor
2MASS photometry; likely this is the reason W0624+6626 was previously overlooked. W0624+6626 is detected on several of the 
POSS plates, POSS1 red, POSS2 red, and POSS2 IR, confirming that it passed very close 
to a background star during the 2MASS epoch, see Figure 2. W0624+6626 shows colors that are red, $W1-W2=0.24\pm0.04$, 
consistent with that of a late M dwarf/early L dwarf \citep{Kirkpatricketal2011}; while the 2MASS $J$ and $H$ band photometry are 
contaminated by a background star.

\begin{figure}
\caption{
Finder charts for W0401+2849, W0404+4127, and W0624+6626, showing the epoch of WISE, 2MASS, POSS2 IR, POSS2 red, and POSS1 red.
For each image the circles indicate the position of the source at the WISE epoch, and the extrapolated 
position of the source at the epoch of that image.
Each image is 3$^{\prime}$ x 3$^{\prime}$, north is up and east is to the left.
}
\begin{center}
\includegraphics[width=5.5in]{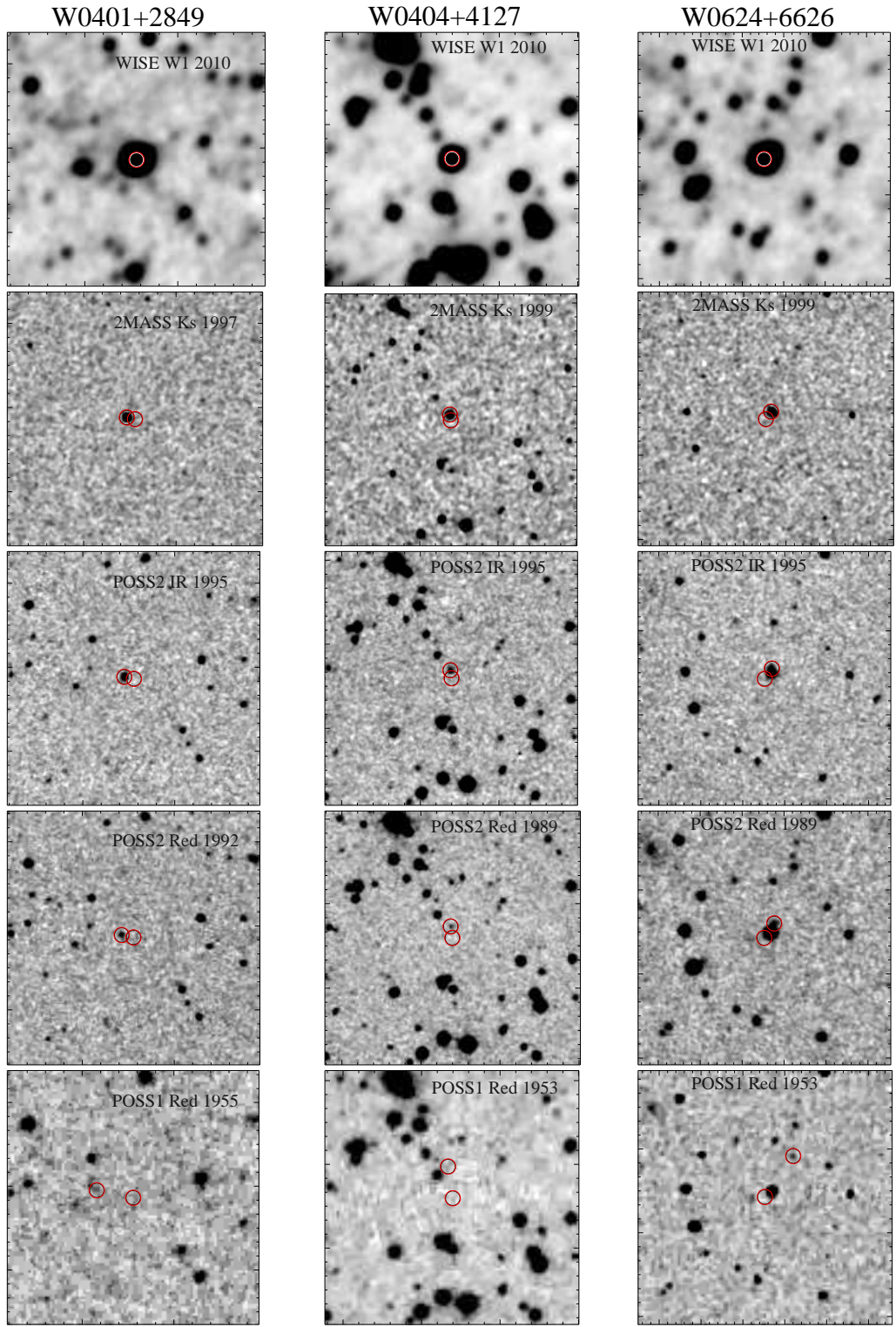}
\end{center}
\end{figure}

\subsection{Proper Motion}
To determine the proper motion of W1405+8350 we used the background stars within 6$^{\prime}$ of the WISE source of interest
as reference to calibrate the positions on the USNO $z$ band image in the 2MASS astrometric reference frame.
We find a proper motion directly from the USNO $z$ and 2MASS astrometry for W1405+8350 to be 
$\mu_{\alpha}$cos($\delta$)$=-0.63\pm0\farcs01$ yr$^{-1}$ and $\mu_{\delta}=0.57\pm0\farcs02$ yr$^{-1}$, 
with total motion $0.85\pm0\farcs02$ yr$^{-1}$.

We determine the proper motions of W0401+2849 and W0404+4127 between the WISE and 2MASS epochs relative 
to the common background stars within 5$^{\prime}$ of the WISE source of interest.
We find a proper motion for W0401+2849 of $\mu_{\alpha}$cos($\delta$)$=-0.48\pm0\farcs01$ yr$^{-1}$
and $\mu_{\delta}=-0.10\pm0\farcs01$ yr$^{-1}$, with total motion $0.49\pm0\farcs02$ yr$^{-1}$.
We find a proper motion for W0404+4127 of $\mu_{\alpha}$cos($\delta$)$=-0.06\pm0\farcs01$ yr$^{-1}$
and $\mu_{\delta}=-0.40\pm0\farcs01$ yr$^{-1}$, with total motion $0.40\pm0\farcs01$ yr$^{-1}$.

Since W0624+6626 is coincident with a background star for the 2MASS epoch, the proper motion needs to be determined using
epochs in which W0624+6626 is resolved from the background star. The only two epochs where this is the case is the 
WISE epoch and POSS1 red epoch, see Figure 2. We determine the proper motion by 
using the background stars within a 5$^{\prime}$ x 5$^{\prime}$ field of view centered on the WISE source of interest as 
reference to calibrate the positions on the POSS1 red image in the WISE astrometric reference frame.
We find a proper motion directly from the WISE and POSS1 red astrometry 
for W0624+6626 to be $\mu_{\alpha}$cos($\delta$)$=0.355\pm0\farcs002$ yr$^{-1}$ and $\mu_{\delta}=-0.513\pm0\farcs003$ yr$^{-1}$, 
with total motion $0.624\pm0\farcs004$ yr$^{-1}$.

\subsection{Transformation}
A transformation between $I_{\rm C}$ and $i$ is determined by using data from Table 2 of \citet{Dahnetal2002} and
Sloan Digital Sky Survey (SDSS) DR8 photometry. We selected L dwarfs from \citet{Dahnetal2002} that have $I_{\rm C}$ photometry
and are not reported as photometrically variable or spectroscopically peculiar. For those sources that had counterparts in
SDSS DR8 \citep{Aiharaetal2011}, we obtained $i$ and $z$ band photometry, resulting in 12 sources with $I_{\rm C}$, $i$,
and $z$ band photometry. One source was removed, the L4.5 dwarf 2M2224-01, due
to conflicting $z$ band photometry from \citet{Dahnetal2002} and SDSS, $\Delta z \approx 0.7$ mag, well outside of the
uncertainty, suggesting erroneous data or variability. Lastly, the matching between \citet{Dahnetal2002} and SDSS DR8
was verified visually. We plot $i$-$z$ vs. $I_{\rm C}$-$i$ and find the best linear least-squares fit to the data,
see Figure 3. The black dashed line shows the best fit line and the red dotted lines show the 1$\sigma$ error in the fit:\\
\\
$I_{\rm C}-i = (-0.0953 \pm -0.0647) + (-0.5010 \pm 0.0349)(i-z)$ for 1.7$ \lesssim i-z \lesssim 3.1$ \\
\\
Using this transformation, we find a value of $i$=20.56$\pm$0.16 for W1405+8350
from $I_{\rm C}$ and $z$ photometry, where the error is from the uncertainty in the photometry and the transformation.
Additional observations of close (bright) mid to late L dwarfs within the coverage of SDSS in the $I_{\rm C}$ band would
further constrain the fit.
The blue dash-dot line is the transformation from \citet{Lupton2005} shown for comparison, which overlaps our
transformation and its uncertainty for mid-L dwarfs (L5-L7), but strongly deviates for early L dwarfs and to a
smaller degree the latest L dwarfs.

Our transformation was originally used to estimate the optical spectral type of W1405+8350, by comparing W1405+8350 to the 
SDSS L dwarfs from \citet{Schmidtetal2010}, using color-color diagrams in the same manner as \citet{CastroGizis2012}.
Our transformation from $I_{\rm C}$ to $i$ for L dwarfs will be useful to the astronomical community, enabling the direct comparison
of L dwarfs with $I_{\rm C}$ and $z$ photometry to the large catalog of L dwarfs from SDSS \citep{Schmidtetal2010}.

\begin{figure}
\caption{
Color-color diagram for L dwarfs observed by \citet{Dahnetal2002}, using $I_{\rm C}$ band photometry from \citet{Dahnetal2002}
and $i$ and $z$ band photometry from SDSS DR8. The black dashed line shows the best linear least-squares fit to the data, with
the red dotted lines showing the 1$\sigma$ error in the fit. The blue dash-dot line is the transformation from \citet{Lupton2005}, which
is consistent with our transformation for mid-L dwarfs but deviates for the latest L dwarfs and strongly deviates for the
early L dwarfs.
}
\begin{center}
\includegraphics[width=6.5in]{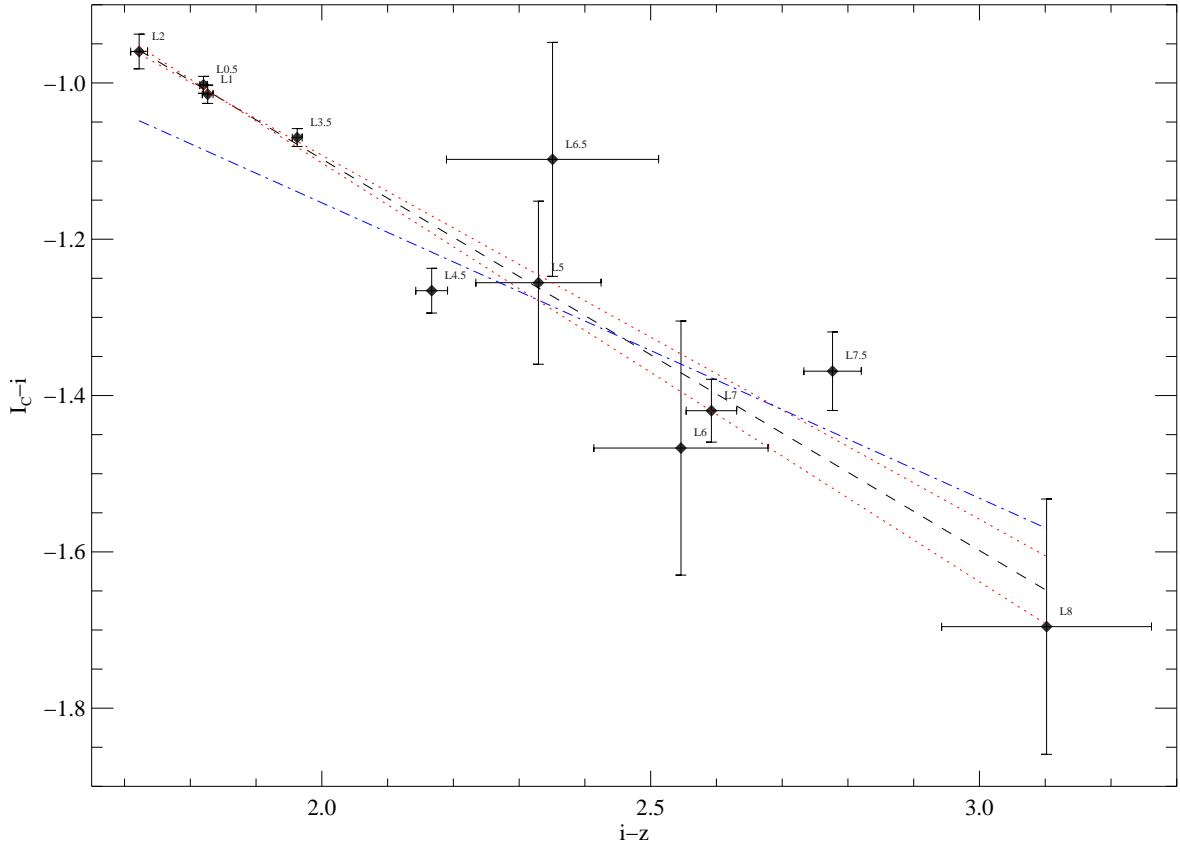}
\end{center}
\end{figure}

\section{OBSERVATIONS}
We obtained optical and near-infrared spectroscopy of all the L dwarfs in order to determine spectral types. See Table 1
for a detailed log of the spectroscopic observations.

\begin{deluxetable}{lccc}
\tablecaption{Optical/Near-infrared Spectroscopy Observation Log}
\tablewidth{0pt}
\tablehead{
\colhead{Object} & \colhead{Telescope/Instr.} & \colhead{Int. (s)} & \colhead{Standard}
}
\startdata
WISE J040137.21+284951.7  & IRTF/SpeX       & 720  & HD 24000 \\
                          & MMT/RCS         & 600  & BD+28 4211 \\
WISE J040418.01+412735.6  & IRTF/SpeX       & 720  & BD+45 981 \\
                          & MMT/RCS         & 600  & BD+28 4211 \\
WISEP J060738.65+242953.4 & IRTF/SpeX       & 960  & HD 46053 \\
                          & MMT/RCS         & 900  & BD+28 4211 \\
WISE J062442.37+662625.6  & IRTF/SpeX       & 540  & HD 38831 \\
                          & MMT/RCS         & 300  & BD+28 4211 \\
WISE J140533.32+835030.5  & APO/TSpec       & 1200 & HD 99966 \\
                          & Keck-II/NIRSPEC & 960  & HD 172864 \\
                          & MMT/RCS         & 1200 & BD+28 4211 \\
\enddata
\end{deluxetable}

\subsection{NIRSPEC with Keck-II}
Moderate-resolution spectra of W1405+8350 were obtained with the Near-Infrared
Spectrometer (NIRSPEC) \citep{McLeanetal1998,McLeanetal2000} at the 10 m W. M. Keck Observatory on 2012 June 7 (UT) and
processed using the IDL software package REDSPEC \citep{McLeanetal2003}. Spectra were obtained in the N3 configuration
with data spanning the $J$ band, 1.14 to 1.36$\mu$m. We used the 42 x $0\farcs57$ slit, giving a resolving power
of R$\sim$2000. Seeing was $\approx1\farcs3$. Two ABBA sequences of 120 s were taken, with a total integration time
of 960 s. The standard HD172864 (A2V) was observed immediately after the science observation at similar airmass for
telluric correction. Wavelength calibration made use of the Ne and Ar calibration lamps. The spectrum has been
normalized at 1.27$\mu$m and corrected for heliocentric motion.

\subsection{TSpec with APO}
W1405+8350 was also observed using TripleSpec \citep{Wilsonetal2004} at the Apache Point Observatory 3.5 m telescope
on UT Date 2 June 2012 under non-photometric conditions. The spectrum was extracted and calibrated using a custom
version of SpeXTool \citep{Vaccaetal2003,Cushingetal2004}. The A-star calibrator was HD 99966. The resulting spectrum
is low signal-to-noise.

\subsection{RCS with MMT}
Optical spectra were obtained on UT Dates 26 and 27 August 2012 using
the MMT Red Channel Spectrograph (RCS) (NOAO Proposal ID: 2012B-0233).
Spectra were obtained with the $1\farcs0$ slit aligned at the parallactic angle. Grating 270 (blazed at 7300 \AA)
with order-blocking filter LP-530 was used to yield wavelength coverage 6200-9800 \AA\ and resolution 12 \AA; strong
telluric water absorption limited the usefulness of the data beyond 9000 \AA.
Conditions were non-photometric.
The data were extracted and calibrated using standard IRAF tasks.

\subsection{SpeX with IRTF}
Low-resolution IRTF SpeX \citep{Rayneretal2003} spectra were
obtained on 2012 February 15 (UT) and processed using SpeXTool \citep{Vaccaetal2003,Cushingetal2004}. Spectra were
obtained in prism mode using the $0\farcs5$ slit aligned at the parallactic angle. The resolution of the corresponding
data spanning 0.7-2.5 $\micron$ was $\lambda$/$\Delta \lambda$ $\approx$ 120. Conditions were non-photometric with
light clouds and the seeing was $>1\arcsec$ at $K$. We obtained individual exposures for each object in an ABBA dither
pattern along the slit. Immediately after the science observations we observed an A0V star for each object at a similar
airmass for telluric corrections and flux calibration. Internal flat field and Ar arc lamp exposures were acquired for
pixel response and wavelength calibration, respectively.

\section{SPECTRAL ANALYSIS OF NEWLY DISCOVERED L DWARFS}
We discuss the spectroscopic observations of each newly discovered L dwarf, followed by their distance 
and physical properties.

\subsection{WISE J140533.32+835030.5}

\subsubsection{Near-infrared \& Optical Spectroscopy}
Figure 4 
shows the moderate-resolution NIRSPEC near-infrared (NIR) $J$ band spectrum of W1405+8350 (black) compared to
NIR standards (red) and a reference dwarf (blue) from
the Brown Dwarf Spectroscopy Survey (BDSS)\footnote{Moderate-resolution L and T dwarf templates were drawn from the
NIRSPEC BDSS Data Archive, http://www.astro.ucla.edu/\textasciitilde{}mclean/BDSSarchive/} \citep{McLeanetal2003} and the
moderate-resolution SpeX library\footnote{Moderate-resolution L dwarf templates were drawn from the IRTF Spectral
Library, http://irtfweb.ifa.hawaii.edu/\textasciitilde{}spex/IRTF\_Spectral\_Library/index.html} \citep{Cushingetal2005}.
While the formal classification standards are also reference dwarfs, we identify this subset of reference dwarfs separately since they
are our primary reference of comparison for spectroscopic classification.
We use the NIR standards of \citet{Kirkpatricketal2010} as anchors to determine the spectral classification.
The K I doublet lines at 1.168, 1.177$\mu$m and 1.243, 1.254$\mu$m are sensitive to surface gravity and thus an indicator of
age \citep{McLeanetal2003,McGovernetal2004}. The equivalent widths of the K I doublets increase as a function of spectral type 
from mid M dwarfs to mid L dwarfs, decreasing for late L dwarfs, and increasing again for the early T dwarfs; with the late L dwarfs 
showing significant scatter, especially the L8 dwarfs. There are FeH bands from 1.1939 
to $1.2389\mu$m, present in mid-M dwarfs, strengthening until about L5, then decaying and are absent in the T dwarfs. 
A doublet of Al I at 1.311 and 1.314$\mu$m is present in mid-M dwarfs but vanishes near the M/L boundary \citep{McLeanetal2003}.
The absence of the Al I doublet indicates the spectral type is later than M, while the presence of the shallow FeH bands indicate 
W1405+8350 must be a late L dwarf or a very early T dwarf. Note the deeper FeH bands for the L$_{\rm n}$7 standard compared to 
that of the shallower features of W1405+8350. The deeper K I doublets of W1405+8350 compared to the L8/L9 dwarfs indicate it is 
an old field dwarf.
From the four deep absorption features of K I we find a radial velocity of $-27\pm14$ km s$^{-1}$. We use the 
goodness-of-fit $\chi^{2}$ as defined in \citet{Burgasser2007a} to compare W1405+8350 to the NIR templates; 
from 1.15 to 1.33$\mu$m, omitting regions with the deep K I absorption lines (1.165 to 1.185$\mu$m and 1.24 to 1.26$\mu$m, 
since they are gravity sensitive) and $>$1.33$\mu$m due to a deviation in wavelength calibration for W1405+8350 relative to 
the templates. The best fit to W1405+8350 is the L$_{\rm n}$9 standard DENIS-P J0255-4700, we adopt a NIR spectral type of L9 
for W1405+8350. 

\begin{figure}
\caption{
Moderate-resolution NIRSPEC NIR $J$ band spectrum of W1405+8350 (black) compared to L/T transition dwarf NIR standards (red)
and a reference dwarf (blue).
From top to bottom;
2MASSI J0103320+193536 observed by \citet{McLeanetal2003} is an L$_{\rm o}$6/L$_{\rm n}$7 \citep{Kirkpatricketal2000,Kirkpatricketal2010},
2MASSW J1632291+190441 observed by \citet{McLeanetal2003} is an L$_{\rm o}$8/L$_{\rm n}$8 \citep{Kirkpatricketal1999,Burgasseretal2006a},
DENIS-P J0255-4700 observed by \citet{Cushingetal2005} is an L$_{\rm o}$8/L$_{\rm n}$9 \citep{Kirkpatricketal2008,Burgasseretal2006a}, and
SDSSp J042348.57-041403.5 observed by \citet{McLeanetal2003} is an L$_{\rm o}$7.5/T$_{\rm n}$0 \citep{Cruzetal2003,Burgasseretal2006a}.
The best match to W1405+8350 is DENIS-P J0255-4700, the L9 NIR standard. All spectra are moderate-resolution, R$\sim$2000,
and are from the BDSS or SpeX, and labeled as such. Prominent spectral features discussed in the text are labeled. The noise for
W1405+8350 is shown on the dashed line at the bottom. All spectra have their flux normalized to the mean of a 0.02 $\micron$ window
centered on 1.27 $\micron$, and offset vertically by integers.
}
\begin{center}
\includegraphics[width=6.5in]{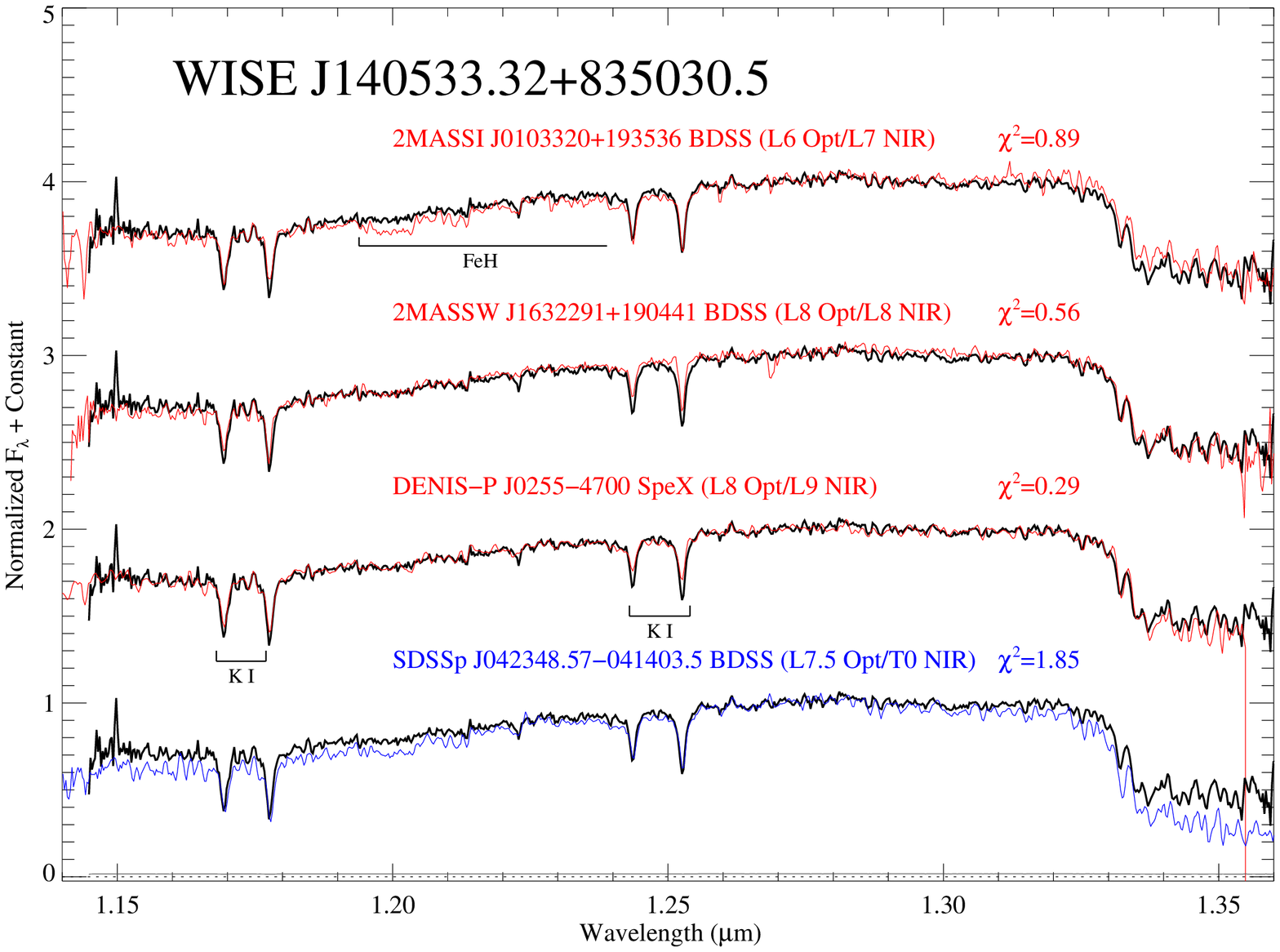}
\end{center}
\end{figure}

Figure 5 shows the smoothed TripleSpec NIR spectrum of W1405+8350 (black) compared to NIR standards (blue, red)
\citep{Kirkpatricketal2010,Burgasseretal2006a} observed with
SpeX\footnote{Low-resolution L and T dwarf templates were drawn from the SpeX Prism Spectral Libraries, http://www.browndwarfs.org/spexprism} 
at low-resolution.
The TripleSpec spectrum is low signal-to-noise, with the $H$ and $K$ bands being shown and 
the $J$ band omitted due to significant noise; the smoothed noise (gray) is shown at the bottom.
The overall appearance of the $H$ band and $K$ band regions is consistent with the L9 NIR spectral
classification.

\begin{figure}
\caption{
Smoothed TripleSpec NIR spectrum of W1405+8350 (black) compared to NIR standards observed with SpeX at low-resolution;
2MASSW J1632291+190441 (blue) observed by \citet{Burgasser2007} is an L$_{\rm o}$8/L$_{\rm n}$8 and
DENIS-P J0255-4700 (red) observed by \citet{Burgasseretal2006a} is an L$_{\rm o}$8/L$_{\rm n}$9.
The smoothed noise (gray) for W1405+8350 is shown at the bottom. 
Although the spectrum of W1405+8350 has considerable noise, the overall appearance of the $H$ and $K$ band regions
are consistent with the L9 NIR spectral classification.
The TripleSpec spectrum is shown from 1.5 to 1.79 $\micron$ and 1.96 to 2.4 $\micron$ in order to 
avoid regions of significant noise.
All spectra have their flux normalized to the mean of a 0.06 $\micron$ window
centered on 1.61 $\micron$.
}
\begin{center}
\includegraphics[width=6.5in]{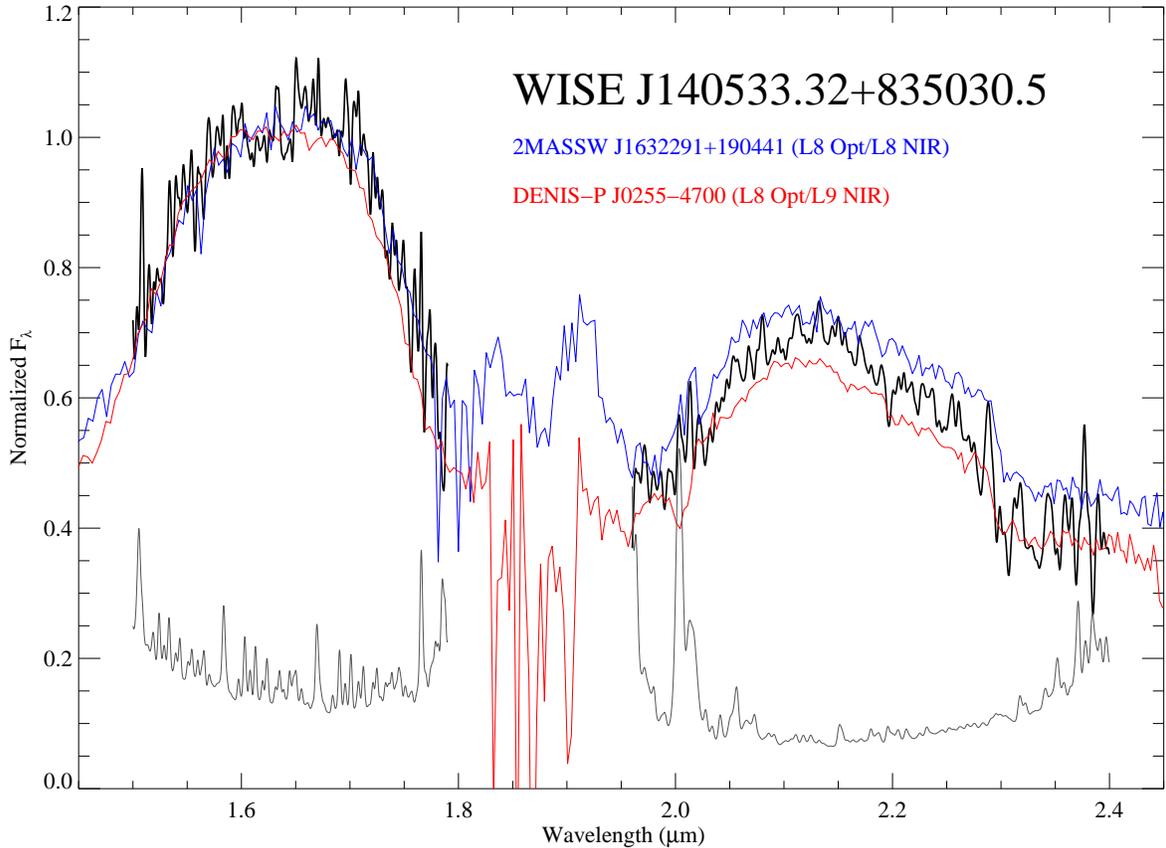}
\end{center}
\end{figure}

Figure 6 shows the optical spectrum for W1405+8350 without telluric corrections. Optical spectral types 
for the L dwarfs were determined by comparing the appearance of the features over the range 6500 \AA\ to 
9000 \AA\ with the standards defined by \citet{Kirkpatricketal1999}. The best fit for W1405+8350
is the L8 optical standard 2MASSW J1632291+190441, we adopt an optical spectral type of L8 for W1405+8350.

\begin{figure}
\caption{
Optical spectra for all of the L dwarfs: WISE J062442.37+662625.6 (L$_{\rm o}$1), WISE J040418.01+412735.6 (L$_{\rm o}$2), WISE J040137.21+284951.7 (L$_{\rm o}$3), 
WISEP J060738.65+242953.4 (L$_{\rm o}$8), and WISE J140533.32+835030.5 (L$_{\rm o}$8).
}
\begin{center}
\includegraphics[width=6.0in]{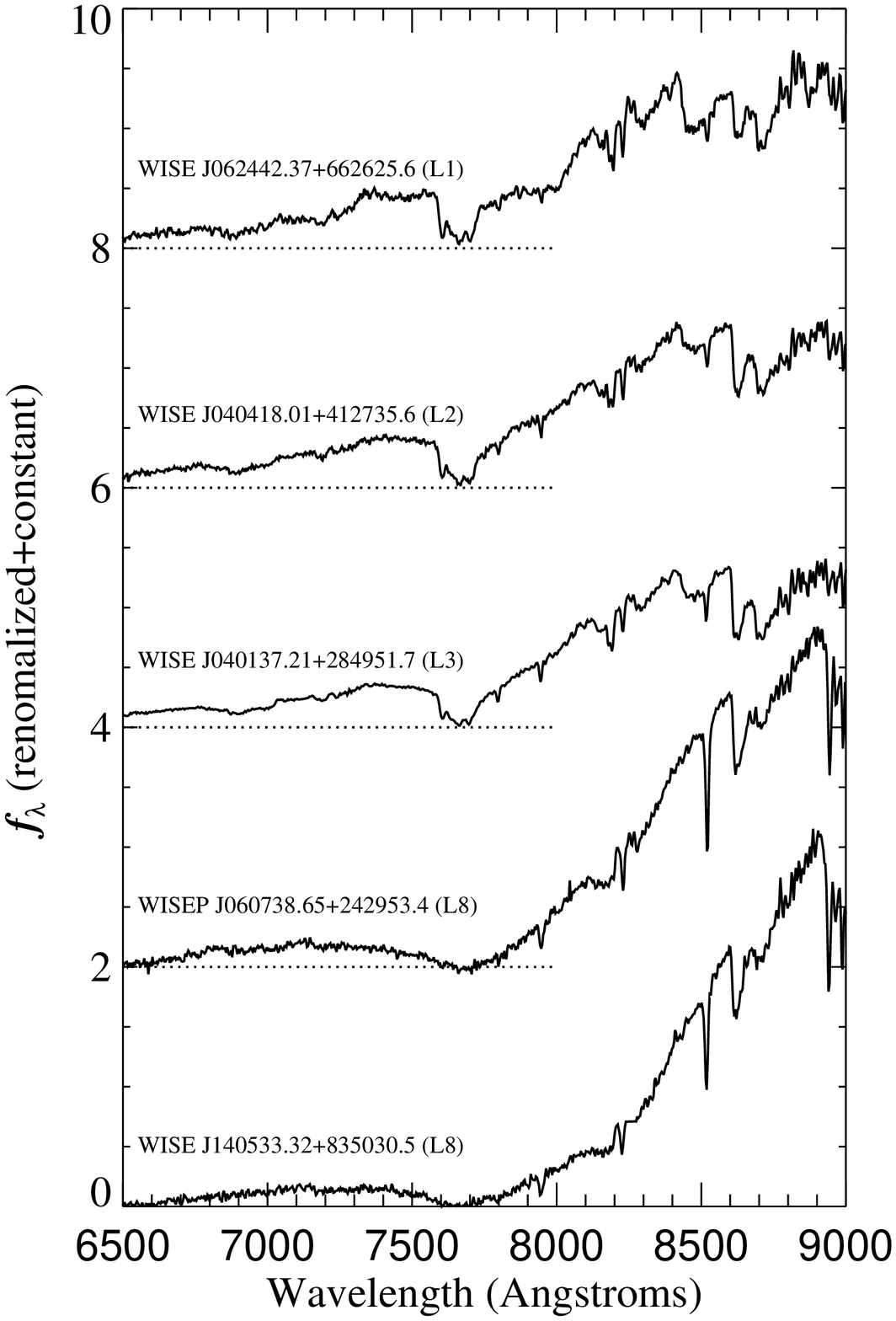}
\end{center}
\end{figure}

\subsubsection{Distance and Physical Properties}
We estimate the distance by using the spectral-type-absolute-magnitude relationships of \citet{Looperetal2008a} for 2MASS 
photometry and the spectral-type-absolute-magnitude relationships of \citet{DupuyLiu2012} for 2MASS and WISE photometry.
We find a distance of $9.5\pm1.3$ pc from 2MASS $J$ photometry, $9.7\pm1.3$ pc from 2MASS $H$ photometry, and
$10.0\pm1.5$ pc from 2MASS $K_{\rm s}$ photometry using the relations from \citet{Looperetal2008a},
$9.8\pm1.8$ pc from 2MASS $J$ photometry, $9.6\pm1.8$ pc from 2MASS $H$ photometry, $10.0\pm1.9$ pc from
2MASS $K_{\rm s}$ photometry, $9.8\pm1.9$ pc from WISE $W1$ photometry, $9.8\pm1.7$ pc from WISE $W2$ photometry,
and $9.4\pm1.9$ pc from WISE $W3$ photometry using the relations from \citet{DupuyLiu2012}. The uncertainty in the 
distance estimates comes from the uncertainty in the photometry and the RMS from the spectral-type-absolute-magnitude 
relationships. The mean of these estimates provides a distance of $9.7\pm1.7$ pc, assuming no binarity.
We note that if we exclude the $J$ band photometry in the distance estimate due to the photometric confusion flag, 
the distance increases by 0.02 pc to 9.8 pc. The proximity of W1405+8350 brings the number of very-late L dwarfs 
within 10 pc to seven, it is a member of the few but growing population of nearby L dwarfs at the L/T transition, 
see Table 2. Trigonometric parallax measurements are needed for a more reliable distance estimate.

\begin{deluxetable}{lccccc}
\tabletypesize{\small}
\rotate
\tablecaption{Very Late L Dwarfs ($\geq$L7) Within 10 Parsecs}
\tablewidth{0pt}
\tablehead{
\colhead{L Dwarf} & \colhead{Spectral Type (Optical/NIR)} & \colhead{Distance (pc)\tablenotemark{a}} & \colhead{$W1-W2$} & \colhead{$W2-W3$} & \colhead{References}
}
\startdata
WISE J104915.57-531906.1A &L8$\pm$1/L7.5 &2.0$\pm$0.15\tablenotemark{b} &0.56$\pm$0.03\tablenotemark{c} & 1.13$\pm$0.03\tablenotemark{c} & 1,[1,2],1\\
DENIS-P J0255-4700 &L8/L9 &4.97$\pm$0.10\tablenotemark{d} &0.55$\pm$0.03 & 1.01$\pm$0.03 & 3,[4,5],6\\
WISEP J060738.65+242953.4 &L8/L9 &7.8$^{+1.4}_{-1.2}$ \tablenotemark{e} &0.59$\pm$0.03 & 0.81$\pm$0.08 & 7,[8,8],7\\
WISEPA J164715.59+563208.2 &$\cdots$/L9 pec (red) &8.6$^{+2.9}_{-1.7}$ \tablenotemark{f} &0.52$\pm$0.03 & 1.03$\pm$0.10 & 9,[$\cdots$,9],9\\
WISEP J180026.60+013453.1 &$\cdots$/L7.5 &8.8$\pm$1.0 &0.47$\pm$0.06 & 1.21$\pm$0.07 & 10,[$\cdots$,10],10\\
2MASS J02572581-3105523 &L8/$\cdots$ &9.7$\pm$1.3 &0.43$\pm$0.03 & 0.99$\pm$0.06 & 4,[4,$\cdots$],11\\
WISE J140533.32+835030.5 &L8/L9 &9.7$\pm$1.7 & 0.58$\pm$0.04 & 1.00$\pm$0.05 & 8,[8,8],8\\
\enddata
\tablerefs{
Discovery, optical and near-infrared spectral type, and distance references: (1) \citet{Luhman2013}; (2) \citet{Burgasseretal2013}; (3) \citet{Martinetal1999}; (4) \citet{Kirkpatricketal2008}; (5) \citet{Burgasseretal2006a}; (6) \citet{Costaetal2006}; (7) \citet{CastroGizis2012}; (8) this work; (9) \citet{Kirkpatricketal2011}; (10) \citet{Gizisetal2011a}; (11) \citet{Looperetal2008b}.}
\tablecomments{WISE colors are from the all-sky data release.}
\tablenotetext{a}{Based on spectrophotometric distance estimate unless otherwise noted.}
\tablenotetext{b}{Based on least-squares fitting of proper and parallactic motion for the astrometry of WISE J104915.57-531906.1AB.}
\tablenotetext{c}{WISE J104915.57-531906.1AB is unresolved in WISE, colors are of the unresolved binary system, the L$_{\rm o}$8$\pm1$/L$_{\rm n}$7.5 \citep{Luhman2013,Burgasseretal2013} primary and the T$_{\rm o}$1.5$\pm2$/T$_{\rm n}$0.5 \citep{Kniazevetal2013,Burgasseretal2013} secondary.}
\tablenotetext{d}{Based on trigonometric parallax.}
\tablenotetext{e}{A preliminary parallax of $139\pm2$ mas ($7.19^{+0.11}_{-0.10}$ pc) was determined based on two seasons of data, consistent with the distance estimate from \citet{CastroGizis2012}.}
\tablenotetext{f}{Based on preliminary astrometric measurements from \citet{Kirkpatricketal2011}, has a spectrophotometric distance estimate of 20.2 pc.}
\end{deluxetable}

Based on the apparent motion and the estimated distance, W1405+8350 has a tangential velocity of $39\pm6$ km s$^{-1}$, 
within range of transverse motions for other L8 dwarfs 
from \citet{Fahertyetal2009}, who quote a median value of 25 km s$^{-1}$ and a dispersion of 19 km s$^{-1}$. This $v_{\rm tan}$ 
is consistent with that expected for a member of the Galactic thin disk \citep{Fahertyetal2009}. Combined with the radial 
velocity from the moderate-resolution $J$ band spectrum of -27$\pm$14 km s$^{-1}$, we find a total velocity of 48$\pm$15 km s$^{-1}$, 
firmly placing it as a member of the galactic thin disk population. Spectral-type-effective-temperature \citep{Looperetal2008a} 
and spectral-type-absolute-bolometric-magnitude \citep{Burgasser2007} relationships give a $T_{\rm eff}=1460\pm90$ K and 
a log $L/L_{\odot}=-4.56\pm0.09$, where the uncertainty in $T_{\rm eff}$ comes from the RMS in the spectral-type-effective-temperature 
relation and the uncertainty in log $L/L_{\odot}$ is from the RMS in the spectral-type-absolute-bolometric-magnitude 
relation. Based on these physical properties, theoretical isochrones from \citet{Baraffeetal2003} give a range
of 0.5 Gyr and 0.03 $M_{\odot}$ to 10 Gyr and 0.072 $M_{\odot}$, placing W1405+8350 in the substellar regime, as are all of the 
latest L dwarfs \citep{Kirkpatrick2005}.
A summary of characteristics for W1405+8350 is found in Table 3.

\begin{deluxetable}{lc}
\tabletypesize{\scriptsize}
\tablecaption{Parameters of WISE J140533.32+835030.5}
\tablewidth{0pt}
\tablehead{
\colhead{Parameters} & \colhead{WISE J140533.32+835030.5}
}
\startdata
USNO R.A. (J2000) &14:05:32.57\\
USNO Decl. (J2000) &+83:50:31.7\\
USNO Epoch &2012.23\\
WISE R.A. (J2000) &14:05:33.32\\
WISE Decl. (J2000) &+83:50:30.6\\
WISE Epoch &2010.25\\
2MASS R.A. (J2000) &14:05:37.29\\
2MASS Decl. (J2000) &+83:50:24.9\\
2MASS Epoch &2000.13\\
$i$ (mag)\tablenotemark{a} &20.56$\pm$0.16\\
USNO $I_{\rm C}$ (mag) &18.93$\pm$0.06\\
USNO $z$ (mag) &17.50$\pm$0.04\\
2MASS $J$ (mag) &14.63$\pm$0.04\\
2MASS $H$ (mag) &13.49$\pm$0.04\\
2MASS $K_{\rm s}$ (mag) &12.85$\pm$0.03\\
WISE $W1$ (mag) &12.08$\pm$0.03\\
WISE $W2$ (mag) &11.50$\pm$0.03\\
WISE $W3$ (mag) &10.50$\pm$0.04\\
WISE $W4$ (mag) &$>$ 9.68\\
$I_{\rm C}-z$ (mag) &1.43$\pm$0.07\\
$I_{\rm C}-J$ (mag) &4.30$\pm$0.07\\
$i-z$ (mag) &3.06$\pm$0.16\\
$i-J$ (mag) &5.93$\pm$0.16\\
$z-J$ (mag) &2.87$\pm$0.06\\
$J-K_{\rm s}$ (mag) &1.78$\pm$0.06\\
$W1-W2$ (mag) &0.58$\pm$0.04\\
$H-W2$ (mag) &1.99$\pm$0.05\\
$K_{\rm s}-W2$ (mag) &1.35$\pm$0.04\\
$\mu_{\alpha}$cos($\delta$) (mas yr$^{-1}$) &-630$\pm$10\\
$\mu_{\delta}$ (mas yr$^{-1}$) &570$\pm$20\\
Total Motion ($\arcsec$ yr$^{-1}$) &0.85$\pm$0.02\\
Spectral Type (Opt/NIR) & L8/L9\\
Distance (pc) &9.7$\pm$1.7\\
${v}_{\rm tan}$ (km s$^{-1}$) &39$\pm$7\\
${v}_{\rm rad}$ (km s$^{-1}$) &-27$\pm$14\\
${v}_{\rm tot}$ (km s$^{-1}$) &48$\pm$16\\
$T_{\rm eff}$ (K) &1460$\pm$90\\
log $L/L_{\odot}$ &-4.56$\pm$0.09\\
Mass at 0.5 Gyr (M$_{\odot}$) &0.030\\
Mass at 10 Gyr (M$_{\odot}$) &0.072\\
\enddata
\tablecomments{WISE data is from the all-sky data release.}
\tablenotetext{a}{Transformed from I$_{\rm C}$ using transformation derived in Section 2.4.}
\end{deluxetable}

Approximately 20\% of L and T dwarfs are resolved as very-low-mass binaries, with the resolved binary fraction of L/T transition 
dwarfs (L7-T3.5) double that of other L and T dwarfs \citep{Burgasseretal2006}. Field binaries are primarily equal 
brightness/mass systems in tightly bound orbits ($<$ 20 AU), where the separation of binary systems peaks 
at $<$ 10 AU \citep{Allen2007,Burgasseretal2007}. A secondary to W1405+8350 of equal or earlier spectral type ($\lesssim$L8) 
would have been detected at $\gtrsim$ 15 AU based on the FWHM ($\approx1\farcs5$) of the USNO $z$ band image. If W1405+8350 
was an unresolved binary system, for example, consisting of two L9 dwarfs, it would push the spectrophotometric distance estimate to 13.8 pc.
The highest resolution imaging/spectroscopy is warranted to search for a companion to W1405+8350. The sensitivity of current 
imaging surveys begins to fall off at separations of $\lesssim3-4$ AU, where there is a model predicted frequency peak of 
binarity \citep{Allen2007}. Close L dwarfs such as W1405+8350, if found to have companions, will play a crucial role in 
extending binaries into this regime.

\subsection{WISE J040137.21+284951.7}
\subsubsection{Near-infrared \& Optical Spectroscopy}
Figure 7
shows the SpeX NIR spectrum of WISE J040137.21+284951.7 (W0401+2849) (black) compared to NIR standards (red)
and a reference dwarf (blue) also observed with SpeX.
We use the goodness-of-fit $\chi^{2}$ as defined in \citet{Burgasser2007a} to compare W0401+2849 to standard/reference
spectra. 
W0401+2849 fits equally well the NIR 
standards Kelu-1 (L$_{\rm n}$2) ($\chi^{2}=0.25$) and 2MASSW J1506544+132106 (L$_{\rm n}$3) ($\chi^{2}=0.25$), 
with an excellent fit ($\chi^{2}\le0.10$) to the reference dwarf DENIS-P J1058.7-1548 (L$_{\rm o}$3/L$_{\rm n}$3) ($\chi^{2}=0.08$). 
We use the spectral indices defined in \citet{Burgasseretal2006a} and the spectral-index relation from
\citet{Burgasser2007} (their Table 3) to obtain spectral indices for W0401+2849 of L3.1$\pm$0.8 (H$_{2}$O-$J$),
L2.6$\pm$1.0 (H$_{2}$O-$H$), and L3.6$\pm$1.1 (CH$_{4}$-$K$), for a mean spectral type of L3.1$\pm$1.0;
see Table 4 for details of the NIR spectral indices of W0401+2849. 
The average disagreement between the spectral-index relation and published classifications is $\sigma$=1.1 subtypes 
for L dwarfs, systematically later for early L dwarfs \citep{Burgasser2007,Burgasseretal2010}.
The spectral-index relation gives a mean spectral type of L2.7$\pm$1.0 and L3.4$\pm$1.0 for the NIR standards 
Kelu-1 (L$_{\rm n}$2) and 2MASSW J1506544+132106 (L$_{\rm n}$3), respectively, too late on average by about half a 
spectral subtype. We adopt a NIR spectral type of L2.5$\pm$0.5 for W0401+2849.

\begin{deluxetable}{lccccc}
\rotate
\tablecaption{L Dwarf Near-Infrared Spectral Indices}
\tablewidth{0pt}
\tablehead{
\colhead{}                        & \colhead{}      & \colhead{}      & \colhead{}      & \multicolumn{2}{c}{Near-Infrared Spectral Type} \\
\colhead{Object} & \colhead{H$_{2}$O-$J$} & \colhead{H$_{2}$O-$H$} & \colhead{CH$_{4}$-$K$} & \colhead{Index\tablenotemark{a}} & \colhead{Direct}
}
\startdata
WISE J040137.21+284951.7 & 0.838 (L3.1) & 0.800 (L2.6) & 1.020 (L3.6) & L3.1$\pm$1.0 & L2.5$\pm$0.5 \\
WISE J040418.01+412735.6 & 0.834 (L3.2) & 0.801 (L2.5) & 1.064 (L1.8) & L2.5$\pm$1.0 & L2 pec (red) \\
WISEP J060738.65+242953.4 & 0.695 (L7.5) & 0.693 (L6.8) & 0.835 (L8.9) & L7.7$\pm$1.0 & L9  \\
WISE J062442.37+662625.6 & 0.904 (L1.3) & 0.836 (L1.1) & 1.059 (L2.0) & L1.5$\pm$1.0 & L1 \\
\enddata
\tablenotetext{a}{Spectral indices defined by \citet{Burgasseretal2006a}, with corresponding spectral type determined from the spectral-index relation in \citet{Burgasser2007}. Final index spectral type is the mean of the individual spectral indices.}
\end{deluxetable}

Figure 6 shows the optical spectrum for W0401+2849 without telluric corrections.
The best fit to W0401+2849 is the L3 optical standard 2MASSW J1146345+223053, we adopt an optical spectral type of L3 for W0401+2849.

\begin{figure}
\caption{
SpeX NIR spectrum of W0401+2849 (black) compared to NIR standards (red) and a reference dwarf (blue).
From top to bottom;
Kelu-1 observed by \citet{Burgasseretal2007a} is an L$_{\rm o}$2/L$_{\rm n}$2 \citep{Kirkpatricketal1999,Kirkpatricketal2010},
DENIS-P J1058.7-1548 observed by \citet{Burgasseretal2010} is an L$_{\rm o}$3/L$_{\rm n}$3 \citep{Kirkpatricketal1999,Knappetal2004}, and
2MASSW J1506544+132106 observed by \citet{Burgasser2007} is an L$_{\rm o}$3/L$_{\rm n}$3 \citep{Gizisetal2000,Kirkpatricketal2010}.
W0401+2849 fits the L2 ($\chi^{2}=0.25$) and L3 ($\chi^{2}=0.25$) NIR standards equally well, 
and is an excellent fit to the reference dwarf DENIS-P J1058.7-1548 ($\chi^{2}=0.08$).
All spectra are low-resolution and obtained with SpeX. Prominent spectral features are labeled. The noise for W0401+2849 is
shown on the dashed line at the bottom. All spectra have their flux normalized to the mean of a 0.04 $\micron$ window centered
on 1.29 $\micron$, and offset vertically to the dashed line.
}
\begin{center}
\includegraphics[width=4.75in]{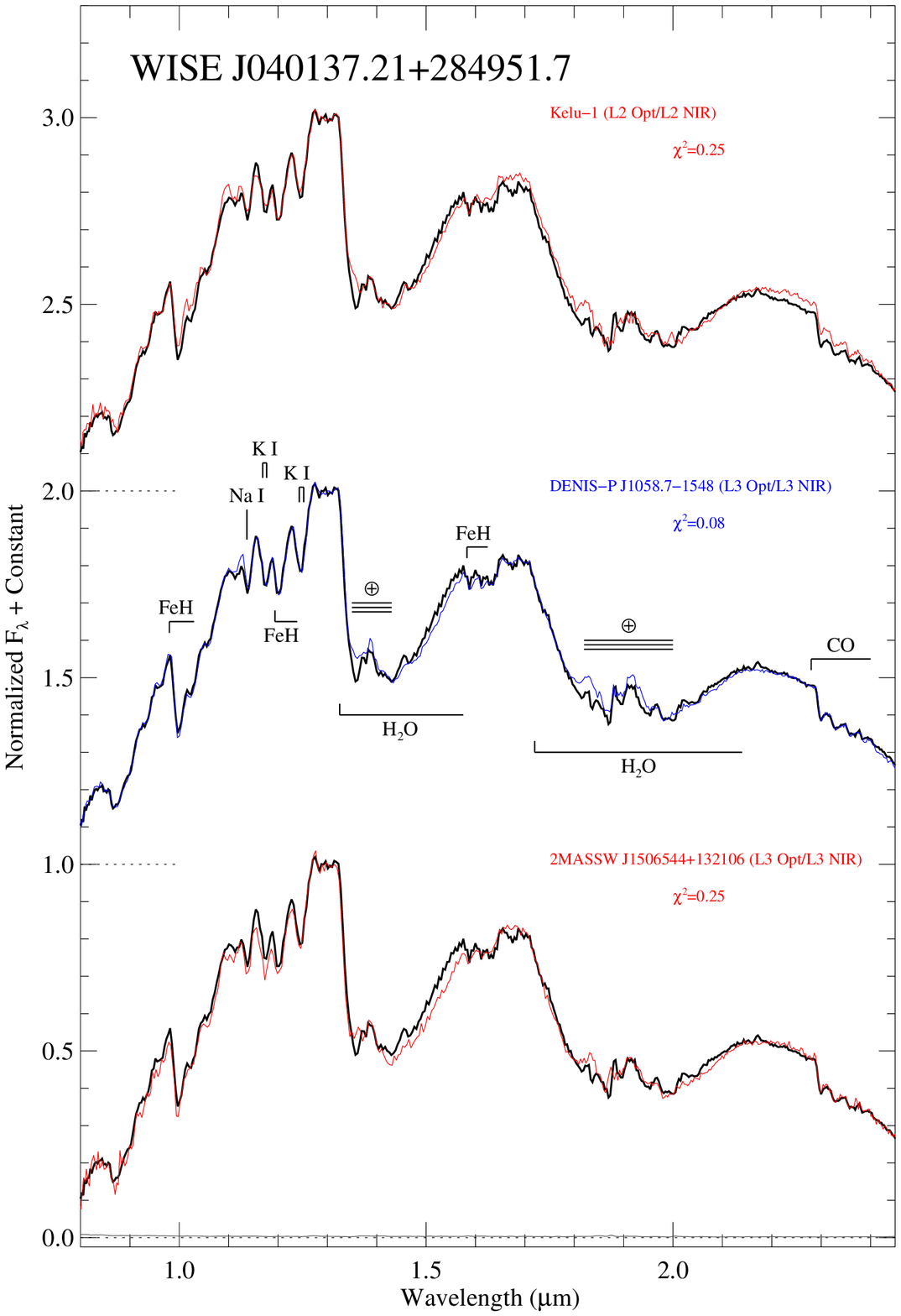}
\end{center}
\end{figure}

\subsubsection{Distance and Physical Properties}
We find a distance of $14.1\pm2.4$ pc from 2MASS and WISE photometry, assuming no binarity. W0401+2849 has a tangential 
velocity of $33\pm6$ km s$^{-1}$, consistent with transverse motions for other L3 dwarfs, with a median value of 30 km s$^{-1}$ 
and a dispersion of 18 km s$^{-1}$. This $v_{\rm tan}$ is consistent with that expected for a member of the Galactic thin disk. 
We find a $T_{\rm eff}=1970\pm90$ K and a log $L/L_{\odot}=-3.93\pm0.09$. Based on these physical properties, theoretical isochrones 
give a range of 0.5 Gyr and 0.05 $M_{\odot}$ to 10 Gyr and 0.075 $M_{\odot}$. A summary of characteristics for W0401+2849 is found 
in Table 5.

\begin{deluxetable}{lccc}
\tablecaption{Parameters of Newly Discovered L Dwarfs}
\tablewidth{0pt}
\tablehead{
\colhead{Parameters} & \colhead{W0401+2849} & \colhead{W0404+4127} & \colhead{W0624+6626}
}
\startdata
WISE R.A. (J2000) &04:01:37.22 & 04:04:18.01 & 06:24:42.38\\
WISE Decl. (J2000) &+28:49:51.7 & +41:27:35.7 & +66:26:25.7\\
WISE Epoch &2010.21 & 2010.22 & 2010.30\\
2MASS R.A. (J2000) &04:01:37.67 & 04:04:18.07 & 06:24:41.72\\
2MASS Decl. (J2000) &+28:49:52.9 & +41:27:39.8 & +66:26:30.9\\
2MASS Epoch &1997.99 & 1999.85 & 1999.91\\
2MASS $J$ (mag) &13.41$\pm$0.02 & 14.15$\pm$0.03 & 13.41$\pm$0.05\tablenotemark{a}\\
2MASS $H$ (mag) &12.43$\pm$0.02 & 13.10$\pm$0.03 & 12.77$\pm$0.05\tablenotemark{a}\\
2MASS $K_{\rm s}$ (mag) &11.82$\pm$0.02 & 12.42$\pm$0.02 & 12.27$\pm$0.03\\
WISE $W1$ (mag) &11.28$\pm$0.02 & 11.85$\pm$0.02 & 12.02$\pm$0.03\\
WISE $W2$ (mag) &11.01$\pm$0.02 & 11.55$\pm$0.02 & 11.79$\pm$0.03\\
WISE $W3$ (mag) &11.01$\pm$0.15 & 11.64$\pm$0.24 & 11.63$\pm$0.17\\
WISE $W4$ (mag) &$>$8.95 & $>$9.03 & $>$9.01\\
$J-K_{\rm s}$ (mag) &1.59$\pm$0.03 & 1.73$\pm$0.04 & 1.14$\pm$0.06\tablenotemark{a}\\
$\mu_{\alpha}$cos($\delta$) (mas yr$^{-1}$) &-480$\pm$10 &  -60$\pm$10 &  355$\pm$2\\
$\mu_{\delta}$ (mas yr$^{-1}$) &-100$\pm$10 & -400$\pm$10 & -513$\pm$3\\
Total Motion ($\arcsec$ yr$^{-1}$) &0.49$\pm$0.02 & 0.40$\pm$0.01 & 0.624$\pm$0.004\\
Spectral Type (Opt/NIR) & L3/L2.5$\pm$0.5 & L2/L2 pec (red) & L1/L1\\
Distance (pc)\tablenotemark{b} &14.1$\pm$2.4 & 23.2$\pm$3.6\tablenotemark{c} & 22.3$\pm$4.1\tablenotemark{d}\\
${v}_{\rm tan}$ (km s$^{-1}$) &          33$\pm$6 &           44$\pm$7 &           67$\pm$12\\
$T_{\rm eff}$ (K)\tablenotemark{e} &1970$\pm$90 & 2100$\pm$90 & 2210$\pm$90\\
log $L/L_{\odot}$\tablenotemark{f} &-3.93$\pm$0.09 & -3.80$\pm$0.09 & -3.69$\pm$0.09\\
Mass at 0.5 Gyr (M$_{\odot}$)\tablenotemark{g} &0.050 & 0.060 & 0.060\\
Mass at 10 Gyr (M$_{\odot}$)\tablenotemark{g} &0.075 & 0.075 & 0.080\\
\enddata
\tablecomments{WISE data is from the all-sky data release.}
\tablenotetext{a}{2MASS `PH\_QUAL' flags of `EEA' for the $JHK_{\rm s}$ bands, respectively, the 2MASS $J$ and $H$ band photometry are contaminated by a background star.}
\tablenotetext{b}{Based on spectral-type-absolute-magnitude relationships of \citet{Looperetal2008a} for 2MASS photometry and the spectral-type-absolute-magnitude relationships of \citet{DupuyLiu2012} for 2MASS and WISE photometry.}
\tablenotetext{c}{Based on 2MASS $J$ band photometry.}
\tablenotetext{d}{Based on 2MASS $K_{\rm s}$ and WISE photometry.}
\tablenotetext{e}{Based on the spectral-type-effective-temperature relationship of \citet{Looperetal2008a}.}
\tablenotetext{f}{Based on the spectral-type-absolute-bolometric-magnitude relationship of \citet{Burgasser2007}.}
\tablenotetext{g}{Based on evolutionary models of \citet{Baraffeetal2003}, and the estimated $T_{\rm eff}$ and log $L/L_{\odot}$.}
\end{deluxetable}

\subsection{WISE J040418.01+412735.6}
\subsubsection{Near-infrared \& Optical Spectroscopy}
Figure 8
shows the SpeX NIR spectrum of WISE J040418.01+412735.6 (W0404+4127) (black) compared to NIR standards (red) and a reference dwarf (blue).
The spectrum of W0404+4127 matches the $J$ band 
of the L2, L3, and L4 NIR standards equally well, while the FeH absorption in the $H$ band identifies it as an early L dwarf (L2-L4).
With excess flux in the $H$ and $K$ bands the spectrum is clearly red, with the slope of the spectrum similar to the L5 NIR standard.
The goodness-of-fit $\chi^{2}$ between W0404+4127 and the (L1-L4) NIR standards was determined in an alternative manner. The spectra were normalized 
to the $J$ band, $H$ band, and $K$ band, and the $\chi^{2}$ test was applied to those respective regions, 
with the $\chi^{2}$ shown being the sum of the three individual $\chi^{2}$ values.
The L2, L3, and L4 NIR standards fit the $J$ band approximately equally well ($\chi^{2}=0.10$, 0.09, 0.10, respectively), 
the L2 fitting the $H$ band best, the L3 fitting the $K$ band best, with the sum of $\chi^{2}$ for the $J$, $H$, and $K$ 
giving the NIR standard Kelu-1 (L$_{\rm n}$2) ($\chi^{2}=0.22$) the best fit to W0404+4127. The spectral-index relation gives L3.2$\pm$0.8 (H$_{2}$O-$J$), 
L2.5$\pm$1.0 (H$_{2}$O-$H$), and L1.8$\pm$1.1 (CH$_{4}$-$K$), for a mean spectral type of L2.5$\pm$1.0; see Table 4 for details 
of the NIR spectral indices of W0404+4127. W0404+4127 has a $J-K_{\rm s}$=1.73$\pm$.04, $\sim$0.2 above the average for 
L2 dwarfs \citep{Fahertyetal2009}, $J-K_{\rm s}$=1.52$\pm$0.20 (where the uncertainty is the 1$\sigma$ standard deviation), typical 
of peculiar red L dwarf discoveries from \citet{Kirkpatricketal2010}. The best fit template to W0404+4127 is 2MASS J08234818+2428577, 
an L3 optical, with no NIR spectral classification found in the literature, it appears to be an overlooked red peculiar L dwarf.

Figure 6 shows the optical spectrum for W0404+4127 without telluric corrections.
The best fit to W0404+4127 is the L2 optical standard Kelu-1, we adopt an optical spectral type of L2 for W0404+4127.

\begin{figure}
\caption{
SpeX NIR spectrum of W0404+4127 (black) compared to NIR standards (red) and a reference dwarf (blue).
From top to bottom;
2MASSW J2130446-084520 observed by \citet{Kirkpatricketal2010} is an L$_{\rm o}$1.5/L$_{\rm n}$1 \citep{Kirkpatricketal2008,Kirkpatricketal2010},
Kelu-1 is an L$_{\rm o}$2/L$_{\rm n}$2, 2MASSW J1506544+132106 is an L$_{\rm o}$3/L$_{\rm n}$3,
2MASS J08234818+2428577 observed by \citet{Burgasseretal2010} is an L$_{\rm o}$3 \citep{Reidetal2008},
2MASS J21580457-1550098 observed by \citet{Kirkpatricketal2010} is an L$_{\rm o}$4:/L$_{\rm n}$4 \citep{Kirkpatricketal2008,Kirkpatricketal2010}, and
SDSS J083506.16+195304.4 observed by \citet{Chiuetal2006} is an L$_{\rm n}$5 \citep{Kirkpatricketal2010}.
The L2 standard ($\chi^{2}=0.22$) is the best statistical fit. For the L1-L4 NIR standards, the goodness-of-fit $\chi^{2}$ was applied 
to the $J$ band, $H$ band, and $K$ band after the spectra were normalized to those respective regions, with the $\chi^{2}$ shown being 
the sum of the three individual $\chi^{2}$ values. 
All spectra are low-resolution and obtained with SpeX. Prominent spectral
features are labeled. The noise for W0404+4127 is shown on the dashed line at the bottom. All spectra have their flux normalized to the mean
of a 0.04 $\micron$ window centered on 1.29 $\micron$, and offset vertically to the dashed line.
}
\begin{center}
\includegraphics[width=4.4in]{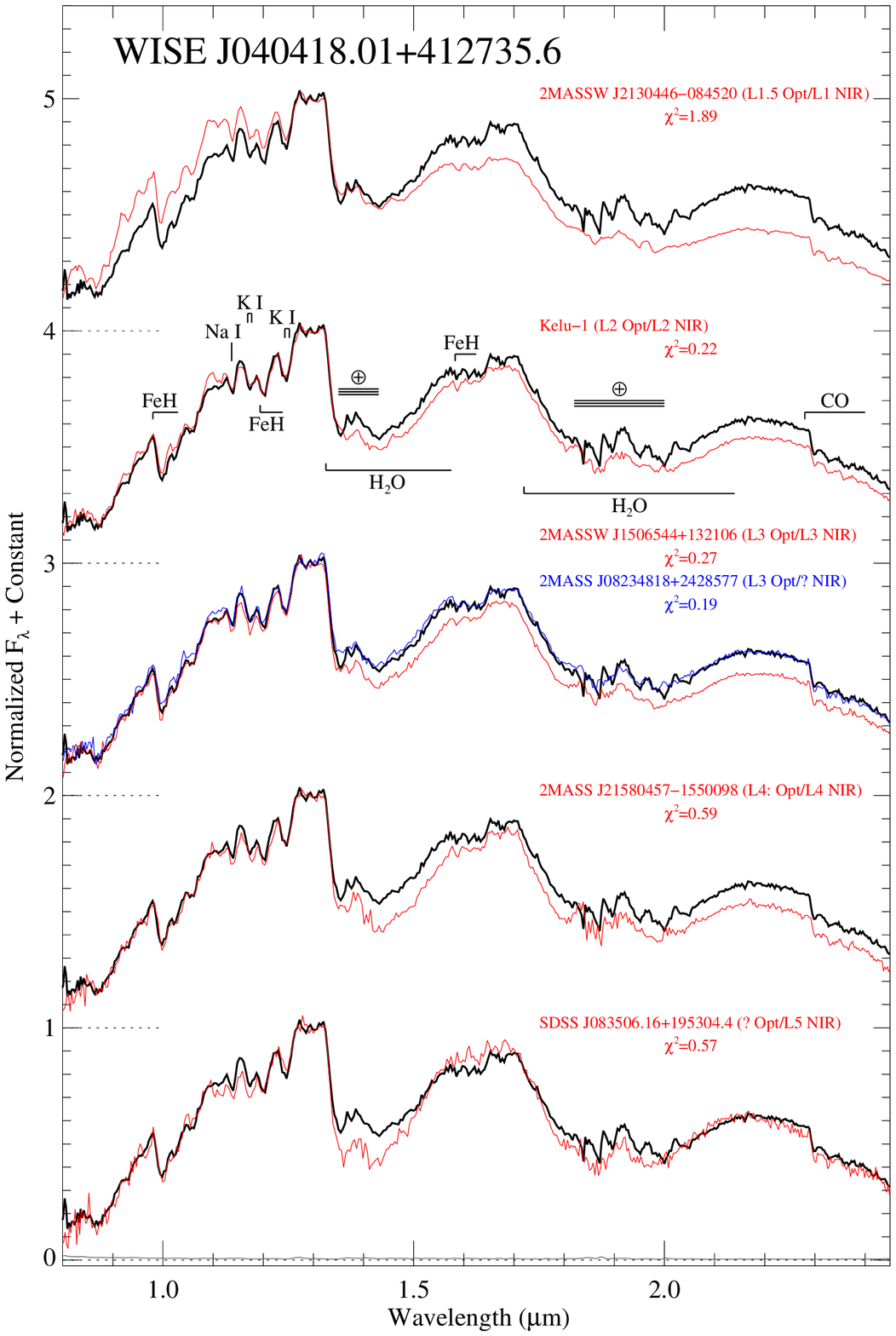}
\end{center}
\end{figure}

\subsubsection{Unusually Red L Dwarf?}
L dwarfs with unusually red SEDs are thought to be the result of thick dust clouds \citep{Looperetal2008b,Cushingetal2008},
where thick dust clouds may be the result of 1. low-gravity (youth) and/or 2. high metallicity; the red spectrum may also be explained 
by 3. unresolved binarity. For more information on peculiar red L dwarfs see \citet{Kirkpatricketal2010,Gizisetal2012,Maceetal2013}.
1. W0404+4127 is likely not a member of the low-gravity (youth) class of red L dwarfs since it lacks any hint of the characteristic
triangular-shaped $H$ band \citep{Kirkpatricketal2006,Kirkpatricketal2010}; with a low-resolution spectrum lacking the ability to 
investigate the individual strengths of gravity sensitive atomic/molecular features. 
W0404+4127 has a tangential velocity of $44\pm7$ km s$^{-1}$, not supportive of youth,
it is at the high end of the 1$\sigma$ dispersion value of transverse motions for other L2 dwarfs from \citet{Fahertyetal2009},
who quote a median value of 26 km s$^{-1}$ and a dispersion of 18 km s$^{-1}$.
2. Metal rich mid-type L dwarfs are expected to have weaker FeH, Na I, K I, and H$_{2}$O absorption, and a redshifted $K$ band peak 
compared to solar-metallicity dwarfs, as predicted by models \citep{Looperetal2008b}. W0404+4127 does not show 
these weaker absorption features compared to the best fit L2 NIR standard, 
it does however show those trends when it is compared with the L3 and L4 NIR standards,
while the $K$ band peak does not appear to be redshifted compared to the L2-L4 NIR standards.
3. Unresolved binarity may explain the red 
spectrum of W0404+4127, where the light combined spectrum would be an intermediate $J$-$K_{\rm s}$ color between an earlier/brighter/bluer 
L dwarf and a later/fainter/redder L dwarf. With the L dwarf sequence peaking at $J$-$K_{\rm s}$=1.81$\pm$0.2 for L7 dwarfs \citep{Fahertyetal2009}, 
it is possible W0404+4127 may be the combined light spectrum of an early L (due to the FeH absorption in the $H$ band) and an $\approx$L7.

\subsubsection{Unresolved Binary?}
We investigate whether the red spectrum of W0404+4127 could be reproduced by the combined light spectrum of two unresolved L dwarfs.
We use the technique of binary spectral template matching following
the procedures from \citet{Burgasser2007a}, using L dwarf NIR standards from \citet{Kirkpatricketal2010}.
The primary and the secondary are scaled based on their flux contribution according to the M$_{K}$-spectral-type 
relation of \citet{Burgasser2007}, with the source and light combined spectra being normalized to the $J$ band.
Figure 9 shows the best fit composite spectrum for W0404+4127, an L2 primary and an L7 secondary, with a fit of $\chi^{2}=0.40$
determined in the standard manner.
The composite fit to the $H$ band ($\chi^{2}=0.09$, normalized to the $J$ band) is almost as good as
the best fitting single Kelu-1 (L$_{\rm n}$2) ($\chi^{2}=0.06$, normalized to the $H$ band), mostly retaining the FeH absorption feature.
However, the fit to the $K$ band is not red enough. Although the composite provides an improvement over the best fit single L dwarf, it does not 
adequately reproduce the red spectrum of W0404+4127.
If the flux contribution from the L7 were marginally more luminous relative to the L2, qualitatively this would increase
the slope of the composite, potentially providing a better fit.
We examine the L2+L7 composite and the effect of including the uncertainty in the M$_{K}$-spectral-type relation of \citet{Burgasser2007},
with an uncertainty in the relation of $\sigma$=0.26 mag.
Allowing the uncertainty to cause M$_{K}$ to be brighter for the L7 at intervals of 0.5$\sigma$ provides an improved fit, 
where the best fit occurs at 2.5$\sigma$ with $\chi^{2}=0.23$. 
The best fit no longer has a deficit in $K$ band flux (except for a slight deficit at longer wavelengths), however, 
it does have a peakier $H$ band ($\chi^{2}=0.09$) and an unrealistic 82\% increase in flux of the $K$ band of the L7.
The L7 2MASSI J0103320+193536 has a $J$-$K_{\rm s}$=2.14$\pm$0.10, $\sim$0.33 above the average for 
L7 dwarfs \citep{Fahertyetal2009}. In order to explain the red spectrum of W0404+4127 as an unresolved binary, the L7 secondary 
may need to be a red peculiar L dwarf itself.

\begin{figure}
\caption{
The spectrum of W0404+4127 (black) shown with the best fit composite spectrum (green) composed of the L2 (red) primary and L7 (blue) 
secondary NIR standards; 2MASSI J0103320+193536 observed by \citet{Cruzetal2004} is an L$_{\rm o}$6/L$_{\rm n}$7 \citep{Kirkpatricketal2000,Kirkpatricketal2010}.
The combined light spectrum fails to adequately reproduce the FeH absorption feature in the $H$ band and lacks the flux in the $K$ band necessary
to explain the red peculiar spectrum of W0404+4127 as an unresolved binary. The source and composite spectra are normalized to the mean of 
a 0.04 $\micron$ window centered on 1.29 $\micron$, the L2 primary and the L7 secondary are scaled based on their flux contribution to the 
composite according to the M$_{K}$-spectral-type relation of \citet{Burgasser2007}.
The noise for W0404+4127 is shown on the dashed line at the bottom.
}
\begin{center}
\includegraphics[width=6.5in]{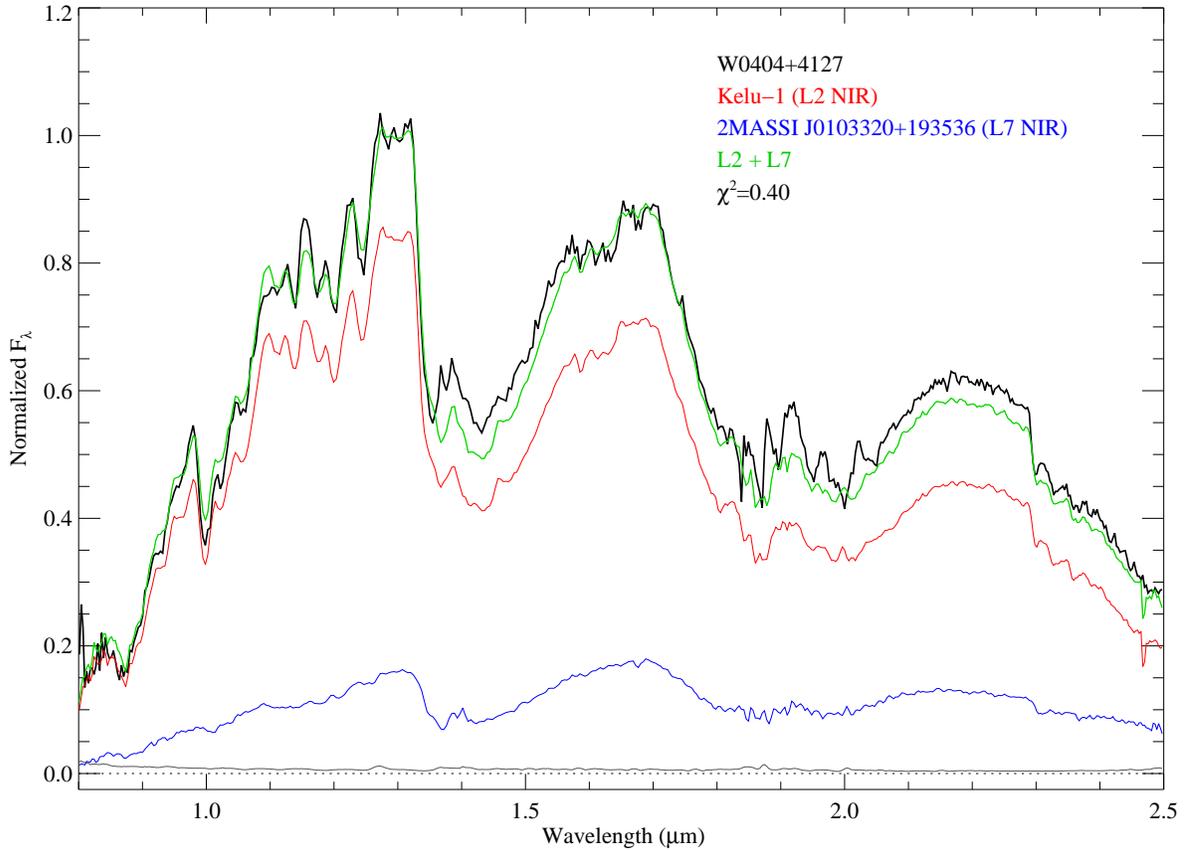}
\end{center}
\end{figure}

\citet{Kirkpatricketal2010} found on average large space velocities from a small sample of red L dwarfs, indicating a kinematically 
older population, in contrast with low gravity (youth) or high metallicity. The kinematics of W0404+4127 are consistent with the 
higher transverse velocities of red L dwarfs from \citet{Kirkpatricketal2010}, suggesting an older age, however, radial velocity 
measurements are needed to confirm a large space velocity. We adopt a NIR spectral type of L2 pec (red) and add W0404+4127 to the 
class of unusually red L dwarfs. 

\subsubsection{Distance and Physical Properties}
We find a distance of $23.2\pm3.6$ pc from 2MASS $J$ photometry, assuming no binarity. We find a $T_{\rm eff}=2100\pm90$ K and a 
log $L/L_{\odot}=-3.80\pm0.09$. Based on these physical properties, theoretical isochrones give a range of 0.5 Gyr and 0.06 $M_{\odot}$ 
to 10 Gyr and 0.075 $M_{\odot}$. A summary of characteristics for W0404+4127 is found in Table 5.

\subsection{WISE J062442.37+662625.6}
\subsubsection{Near-infrared \& Optical Spectroscopy}
Figure 10
shows the SpeX NIR spectrum of WISE J062442.37+662625.6 (W0624+6626) (black) compared to NIR standards (red) and a reference dwarf (blue).
W0624+6626 best fits the L1 NIR standard 2MASSW J2130446-084520,
with a very slight lack of flux in the H and K band compared to 2MASSW J2130446-084520. The best fitting template to W0624+6626 
is DENIS-P J170548.38-051645.7 ($\chi^{2}=0.10$), 
classified as an L$_{\rm n}$4 by \citet{Kendalletal2004}, in contradiction to the spectrum shown here,
and given a NIR SpeX spectral type of L1.0 by \citet{Burgasseretal2010}. 
The spectral-index relation gives L1.3$\pm$0.8 (H$_{2}$O-$J$), L1.1$\pm$1.0 (H$_{2}$O-$H$), 
and L2.0$\pm$1.1 (CH$_{4}$-$K$), for a mean spectral type of L1.5$\pm$1.0, within uncertainty of the direct fit; see Table 4 for 
details of the NIR spectral indices of W0624+6626. The spectra-index relation gives a mean spectral type 
of L1.0$\pm$1.0, L1.4$\pm$1.0, and L2.7$\pm$1.0 for the NIR standards 2MASS J03454316+2540233 (L$_{\rm n}$0), 
2MASSW J2130446-084520 (L$_{\rm n}$1), and Kelu-1 (L$_{\rm n}$2), respectively, too late on average by about half a spectral subtype. 
We adopt a NIR spectral type of L1 for W0624+6626. 

Figure 6 shows the optical spectrum for W0624+6626 without telluric corrections.
The best fit to W0624+6626 is the L1 optical standard 2MASSW J1439284+192915, we adopt an optical spectral type of L1 for W0624+6626.

\begin{figure}
\caption{
SpeX NIR spectrum of W0624+6626 (black) compared to NIR standards (red) and a reference dwarf (blue).
From top to bottom;
2MASS J03454316+2540233 observed by \citet{BurgasserMcElwain2006} is an L$_{\rm o}$0/L$_{\rm n}$0 \citep{Kirkpatricketal1999,Kirkpatricketal2010},
DENIS-P J170548.38-051645.7 observed by \citet{Burgasseretal2010} is an L$_{\rm o}$0.5/L$_{\rm n}$4 \citep{Reidetal2006,Kendalletal2004},
2MASSW J2130446-084520 is an L$_{\rm o}$1.5/L$_{\rm n}$1, and Kelu-1 is an L$_{\rm o}$2/L$_{\rm n}$2.
W0624+6626 best fits the L1 NIR standard 2MASSW J2130446-084520. All spectra are low-resolution and obtained with SpeX. Prominent spectral
features are labeled. The noise for W0624+6626 is shown on the dashed line at the bottom. All spectra have their flux normalized to the mean
of a 0.04 $\micron$ window centered on 1.29 $\micron$, and offset vertically to the dashed line.
}
\begin{center}
\includegraphics[width=5.0in]{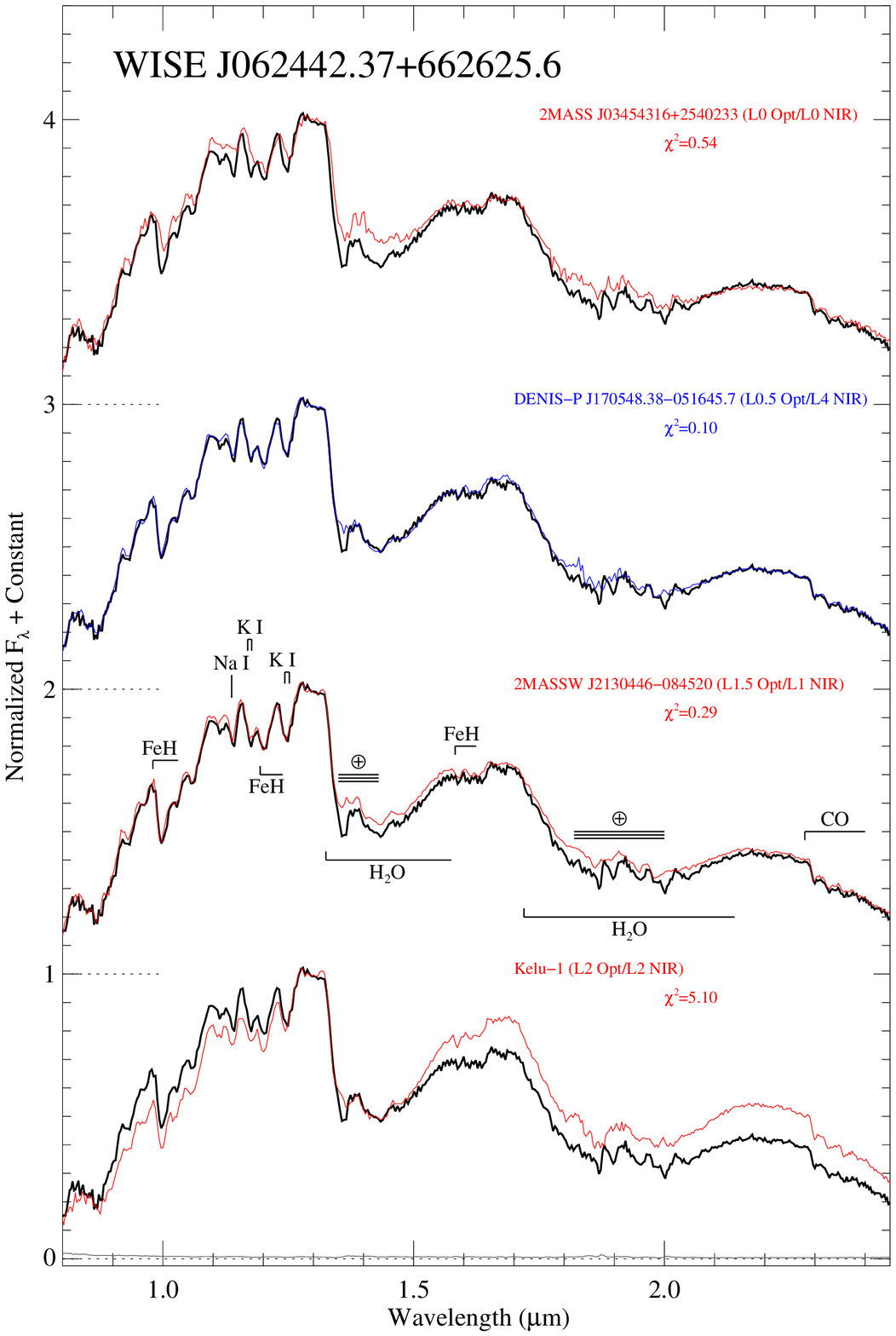}
\end{center}
\end{figure}

\subsubsection{Distance and Physical Properties}
We find a distance of $22.3\pm4.1$ pc from 2MASS $K_{\rm s}$ and WISE photometry, assuming no binarity. W0624+6626 has a tangential 
velocity of $67\pm12$ km s$^{-1}$, above the 1$\sigma$ dispersion value of transverse motions for other L1 dwarfs, with a median 
value of 32 km s$^{-1}$ and a dispersion of 23 km s$^{-1}$. This $v_{\rm tan}$ is consistent with that expected for a member of 
the Galactic thin disk. We find a $T_{\rm eff}=2210\pm90$ K and a log $L/L_{\odot}=-3.69\pm0.09$. Based on these physical properties, 
theoretical isochrones give a range of 0.5 Gyr and 0.06 $M_{\odot}$ to 10 Gyr and 0.08 $M_{\odot}$. A summary of characteristics
for W0624+6626 is found in Table 5.

\section{SPECTROSCOPIC FOLLOW-UP OF WISEP J060738.65+242953.4}
In this section we present follow-up spectroscopy of a previously discovered fifth 
object WISEP J060738.65+242953.4 (W0607+2429), observed during the same night with SpeX at the IRTF
as three of the newly discovered L dwarfs. W0607+2429 was discovered by \citet{CastroGizis2012} as a part 
of the search for high proper motion objects between 2MASS and WISE. They estimate the optical spectral type 
to be L8 based on color-color diagrams using 2MASS and SDSS photometry, with a nearby spectrophotometric 
distance estimate of 7.8 pc.

\subsection{Near-infrared \& Optical Spectroscopy}
Figure 11 shows the SpeX NIR spectrum of W0607+2429 (black) compared to L/T transition dwarf NIR standards (red) 
and a reference dwarf (blue). W0607+2429 is a very good fit to the L9 NIR standard DENIS-P J0255-4700.
The spectral-index relation gives L7.5$\pm$0.8 (H$_{2}$O-$J$),
L6.8$\pm$1.0 (H$_{2}$O-$H$), and L8.9$\pm$1.1 (CH$_{4}$-$K$), for a mean spectral type of L7.7$\pm$1.0. We note that the
weak absorption at the 1.15 $\micron$ H$_{2}$O bands gives a slightly earlier H$_{2}$O-$J$ spectral-index estimate.
The index spectral type is slightly earlier than the direct spectral type of L9, 
but is consistent within 2$\sigma$ of the uncertainty.
The spectral-index relation produces early spectral types compared to published classifications for late
L dwarfs \citep{Burgasser2007,Burgasseretal2010}.
The spectral-index relation gives a mean spectral type
of L7.1$\pm$1.0 for 2MASSW J1632291+190441 (L$_{\rm o}$8/L$_{\rm n}$8), L8.0$\pm$1.0 for DENIS-P J0255-4700
(L$_{\rm o}$8/L$_{\rm n}$9), L8.7$\pm$1.0 for 2MASSI J0328426+230205 (L$_{\rm o}$8/L$_{\rm n}$9.5), and
L9.5$\pm$0.8 for SDSS J120747.17+024424.8 (L$_{\rm o}$8/T$_{\rm n}$0), all within the uncertainty of their published
NIR spectral types but systematically earlier by a mean subtype of 0.8. Table 4
provides details of the NIR spectral indices of W0607+2429.

\begin{figure}
\caption{
SpeX NIR spectrum of W0607+2429 (black) compared to L/T transition dwarf NIR standards (red) and a reference dwarf (blue).
From top to bottom;
2MASSW J1632291+190441 is an L$_{\rm o}$8/L$_{\rm n}$8,
DENIS-P J0255-4700 is an L$_{\rm o}$8/L$_{\rm n}$9,
2MASSI J0328426+230205 observed by \citet{Burgasseretal2008} is an L$_{\rm o}$8/L$_{\rm n}$9.5 \citep{Kirkpatricketal2000,Knappetal2004},
and SDSS J120747.17+024424.8 observed by \citet{Looperetal2007a} is an L$_{\rm o}$8/T$_{\rm n}$0 \citep{Hawleyetal2002,Burgasseretal2006a}.
The best match to W0607+2429 is DENIS-P J0255-4700, the L9 NIR standard. Ambiguous CH$_{4}$ absorption
at 1.6 $\micron$ is present in the $H$ band. The vertical dashed lines show the 2$\nu_{2}+\nu_{3}$ band of CH$_{4}$
at 1.63 $\micron$ and the 2$\nu_{3}$ band of CH$_{4}$ at 1.67 $\micron$ \citep{McLeanetal2003}.
All spectra are low-resolution and obtained with SpeX. Prominent spectral features are labeled. The noise for W0607+2429 is
shown on the dashed line at the bottom. All spectra have their flux normalized to the mean of a 0.04 $\micron$ window centered
on 1.29 $\micron$, and offset vertically to the dashed line.
}
\begin{center}
\includegraphics[width=4.5in]{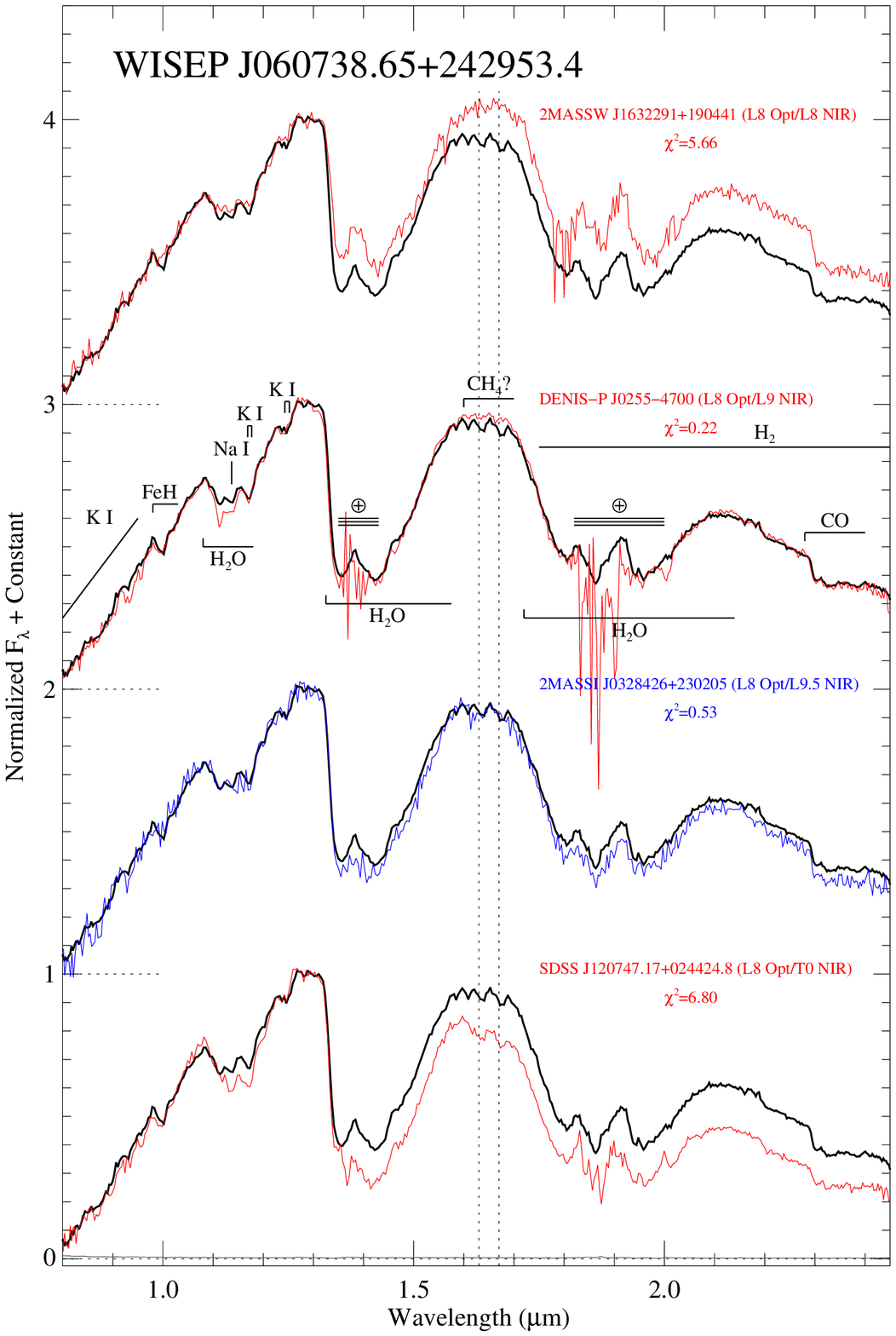}
\end{center}
\end{figure}

The spectrum of W0607+2429 shows weaker absorption at the 1.15 $\micron$ H$_{2}$O bands compared to DENIS-P J0255-4700,
but is similar to 2MASSW J1632291+190441 and 2MASSI J0328426+230205; it may be that DENIS-P J0255-4700 has unusually
strong 1.15 $\micron$ H$_{2}$O bands and W0607+2429 is normal. W0607+2429, similar to DENIS-P J0255-4700 and
2MASSI J0328426+230205, show weak signs of the 2.2 $\micron$ CH$_{4}$ band that is typical of the latest-type L
dwarfs \citep{Burgasser2007a}. Interestingly, W0607+2429 has `ambiguous' CH$_{4}$ absorption at 1.6 $\micron$ in
the $H$ band compared to DENIS-P J0255-4700, with a downward sloping peak compared to the more rounded peak of
DENIS-P J0255-4700. The CH$_{4}$ absorption in the $H$ band is rather similar to 2MASSI J0328426+230205 (L$_{\rm o}$8/L$_{\rm n}$9.5),
but not as deep as SDSS J120747.17+024424.8 (L$_{\rm o}$8/T$_{\rm n}$0). We consider the unambiguous detection of CH$_{4}$
in the $H$ band to be that of the T0 standard \citep{Burgasseretal2006a} SDSS J120747.17+024424.8, where the unambiguous
detection at low resolution of CH$_{4}$ at the H and K bands is the defining characteristic of the T dwarf spectral
class \citep{Kirkpatrick2005}. The methane index in the $H$ band is CH$_{4}$-$H$=1.00 for W0607+2429, 1.08, 1.03, 1.01, and 0.95
for 2MASSW J1632291+190441 (L$_{\rm n}$8), DENIS-P J0255-4700 (L$_{\rm n}$9), 2MASSI J0328426+230205 (L$_{\rm n}$9.5),
and SDSS J120747.17+024424.8 (T$_{\rm n}$0), respectively; with CH$_{4}$-$H>$0.97 indicating that CH$_{4}$ absorption is
not `present' in the $H$ band \citep{Burgasser2007}. We note that 2MASSI J0328426+230205 is suspected of being an unresolved
binary \citep{Burgasseretal2010}, likely due to the same `ambiguous' CH$_{4}$ absorption found in the $H$ band of W0607+2429.

Figure 6 shows the optical spectrum for W0607+2429 without telluric corrections.
The best fit to W0607+2429 is the L8 optical standard 2MASSW J1632291+190441, we adopt an optical spectral type of L8 for W0607+2429.

\subsection{Unresolved Binary?}
Approximately 20\% of L and T dwarfs are resolved as very-low-mass binaries, with the resolved binary fraction of L/T
transition dwarfs (L7-T3.5) double that of other L and T dwarfs \citep{Burgasseretal2006}. Numerous L dwarfs with spectra
exhibiting features of methane in the $H$ band have been shown to be unresolved L/T binaries based on the convention of
empirical binary templates \citep{Burgasseretal2010,Burgasser2007a}. The `ambiguous' CH$_{4}$ absorption in the $H$ band
is suggestive that W0607+2429 could be an unresolved L/T binary. The combined light spectrum from a late L dwarf and an
early T dwarf can give a direct/index spectral type of L9 \citep{Looperetal2007a,Burgasser2007}. Based on synthetic spectral
binary templates from \citet{Looperetal2007a}, an unresolved binary of an L8+T0 (NIR) would give a direct comparison spectral
type of L9, and an L8+T1, L9+T0, and L9+T1 would yield an L9.5 spectral type.

We investigate whether the CH$_{4}$ absorption in the $H$ band of W0607+2429 could be reproduced by the combined light spectrum 
of a primary L dwarf and a secondary T dwarf. We construct composite spectra in the same manner as for W0404+4127, however, 
here we use L and T NIR standards from \citet{Kirkpatricketal2010,Burgasseretal2006a}; with the exception of the T1 NIR standard, 
where we use SDSS J015141.69+124429.6 from \citet{Burgasser2007}.
Figure 12 shows the best fit composite spectrum, the only binary template with $\chi^{2}$ $<$ 0.50, an L8+T0. 
The combined light spectrum of the L8+T0 ($\chi^{2}=0.34$) reproduces 
the CH$_{4}$ absorption feature in the $H$ band, albeit slightly deeper, but does not provide a better fit than the 
single L9 standard ($\chi^{2}=0.22$), with the flux in the $H$ and $K$ band too weak.
Combinations of an L8 primary 
and later secondary T dwarfs result in CH$_{4}$ absorption that is too deep and flux in the $H$ and $K$ band increasingly 
weaker. Likewise, combinations of an L7 primary and T dwarf secondaries do not reproduce the CH$_{4}$ in the $H$ band 
and have excess flux in the $K$ band. 
For the L8+T0 composite, if the flux contribution from the T0 were lowered relative to the L8, qualitatively this would increase 
the $H$ and $K$ band flux, potentially providing a better fit.
We examine the L8+T0 composite and the effect of including the uncertainty in the M$_{K}$-spectral-type relation of \citet{Burgasser2007}.
Allowing the uncertainty to cause M$_{K}$ to be fainter for the T0 from 0$\sigma$ to 1$\sigma$ at intervals of 0.25$\sigma$
results in an increase in the $H$ and $K$ band flux and an improving fit from $\chi^{2}=0.34$ to $\chi^{2}=0.19$. However, in the 
process the $J$ band of the T0 is reduced relative to the L8 such that at 1$\sigma$ the $J$ band peak of the L8 and the T0 are equivalent, 
the characteristic $J$ band bump between late L and early T dwarfs has vanished, and even slightly reversed.
If an uncertainty of 0.75$\sigma$ is allowed in the M$_{K}$-spectral-type relation causing the T0 to be fainter relative
to the L8, then the L8+T0 composite provides a fit of $\chi^{2}=0.21$, slightly better than the single L9 NIR standard ($\chi^{2}=0.22$).
This improved fit comes with a $J$ band bump that is almost absent and a 16\% reduction in flux of the $K$ band of the T0.
The best fit occurs at 1.25$\sigma$ and 1.5$\sigma$, with $\chi^{2}=0.18$, where the $J$ band bump has reversed.
The binary templates do not lend credence to W0607+2429 being an 
unresolved L/T transition binary, nor do they exclude it. It may be that the ambiguous appearance of CH$_{4}$ absorption 
is a natural phenomenon, in this case the onset of CH$_{4}$ in the $H$ band as early as an L9 rather than an L9.5. The highest 
resolution imaging and radial velocity measurements are warranted to further investigate the potential binarity of W0607+2429.

\begin{figure}
\caption{
The spectrum of W0607+2429 (black) shown with the best fit composite spectrum (green) composed of the L8 (red) primary and the T0 (blue)
secondary NIR standards. The combined light spectrum of the L8+T0 ($\chi^{2}=0.34$) reproduces the CH$_{4}$ absorption feature in the $H$ band,
but does not provide a better fit than the single L9 standard ($\chi^{2}=0.22$), with the flux in the $H$ and $K$ band too weak. The source and
composite spectra are normalized to the mean of a 0.04 $\micron$ window centered on 1.29 $\micron$, the L8 primary and the
T0 secondary are scaled based on their flux contribution to the composite according to the M$_{K}$-spectral-type relation
of \citet{Burgasser2007}. The noise for W0607+2429 is shown on the
dashed line at the bottom.
}
\begin{center}
\includegraphics[width=6.5in]{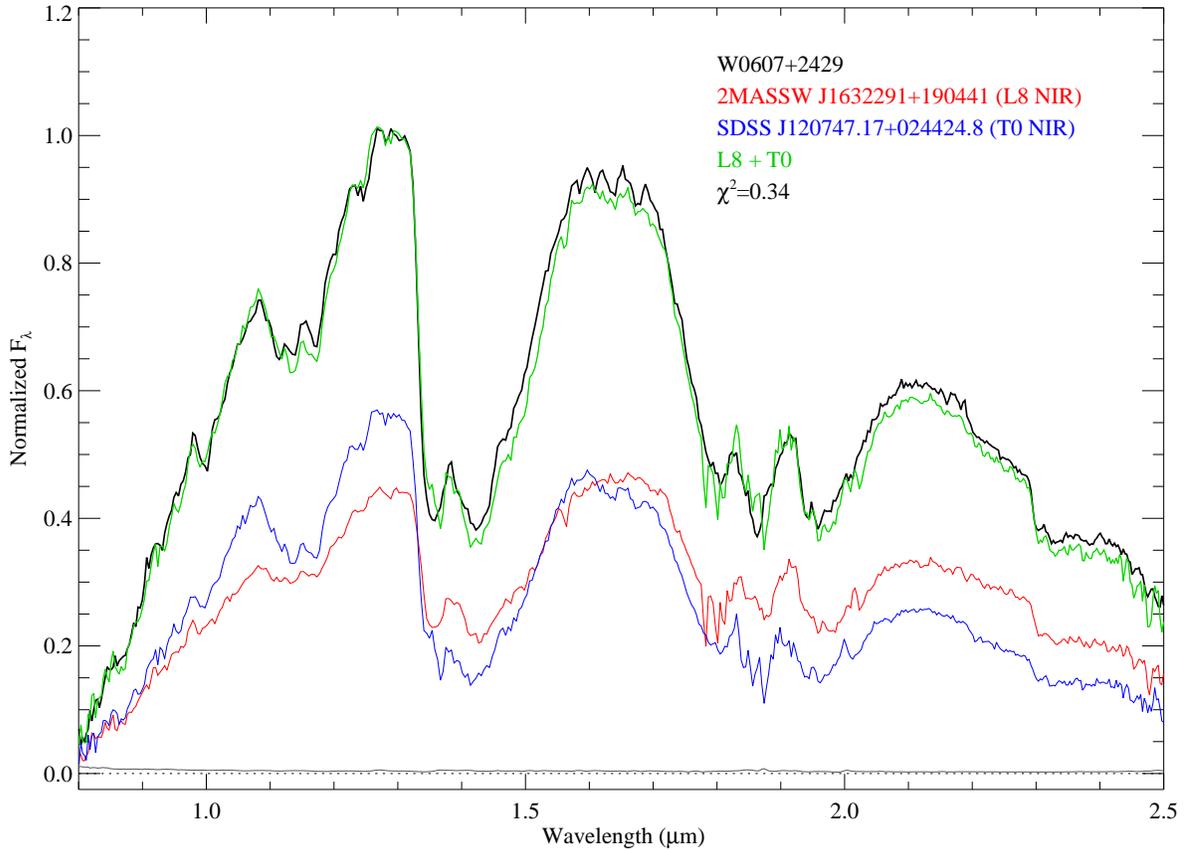}
\end{center}
\end{figure}

The L9 NIR standard DENIS-P J0255-4700 shows weak CH$_{4}$ absorption in the $H$ band in the medium resolution SpeX
spectrum \citep{Cushingetal2005}, but no `distinct' CH$_{4}$ absorption in the low resolution SpeX
spectrum \citep{Burgasseretal2006a}. \citet{Cushingetal2005} argued the case for whether DENIS-P J0255-4700 should be
classified as a T dwarf based on the definition of \citet{Geballeetal2002}, who define the boundary between the
L/T transition as the earliest appearance of methane absorption in both the $H$ and $K$ bands.
\citet{Burgasseretal2006a} classified DENIS-P J0255-4700 as an L9 based on an analogous definition by \citet{Burgasseretal2006a},
that the T dwarf sequence begins with the detection of CH$_{4}$ absorption in the $H$ band at `low' resolution. Dwarfs
classified as Late L that show `distinct' CH$_{4}$ absorption at low resolution akin to W0607+2429 in the $H$ band, such
as 2MASSI J0328426+230205, are generally classified as L$_{\rm n}$9.5. W0607+2429 does not fit the
spectral energy distribution (SED) of L9.5 dwarfs well with excess flux in the $H$ and $K$ bands comparatively;
compared to all templates in the SpeX library the best fit to W0607+2429 being the L9 NIR standard DENIS-P J0255-4700.
By an analogous argument to DENIS-P J0255-4700, one could argue
for the classification of W0607+2429 as a T dwarf based on the presence of `distinct' CH$_{4}$ absorption in the $H$ band at
low resolution, meeting the definition of \citet{Burgasseretal2006a} for a T dwarf. However, the overall SED of W0607+2429
is clearly earlier than T0, in excellent agreement with DENIS-P J0255-4700. W0607+2429 shows the earliest onset of CH$_{4}$
in the $H$ band compared to all of the L9 dwarfs in the SpeX library that have an overall SED consistent with the
L9 NIR standard, DENIS-P J0255-4700; with several objects classified as L9 showing CH$_{4}$ in the $H$ band whose $H$
and $K$ bands are weaker than the L9 NIR standard and not consistent with the overall SED. W0607+2429 may be the catalyst
for a modification to what is defined as a T dwarf if it is indeed a single object, with the onset of `distinct' CH$_{4}$
in the $H$ band as early as L9; and may help to shape our understanding of the L/T transition.

Based on the direct fit and the systematically early index spectral classification, we adopt a NIR spectral type of L9
for W0607+2429.
The adopted optical and NIR spectral type
confirm the optical spectral type estimate from \citet{CastroGizis2012},
illustrating the utility and accuracy of color-color diagrams in estimating the spectral type of normal colored very late L dwarfs.
Discrepant optical/NIR spectral types and CH$_{4}$ absorption features in the $H$ band are
characteristics of unresolved L/T binary systems \citep{Cruzetal2004,Burgasseretal2005,Burgasser2007a,Burgasseretal2010}.
The complementary optical and NIR spectral types of W0607+2429 provide support that it is a single object.

\section{CONCLUSIONS}
We have discovered four high proper motion L dwarfs within 25 pc. WISE J140533.32+835030.5 is an L dwarf at the L/T transition 
within 10 pc, with an optical spectral type of L8, a near-infrared spectral type of L9 from moderate-resolution $J$ band spectroscopy, 
and a proper motion of $0.85\pm0\farcs02$ yr$^{-1}$. We find a distance of $9.7\pm1.7$ pc, increasing the number of L dwarfs at the L/T 
transition within 10 pc from six to seven. WISE J040137.21+284951.7, WISE J040418.01+412735.6, 
and WISE J062442.37+662625.6 are all early L dwarfs within 25 pc. WISE J040418.01+412735.6 is a member of the class of unusually red L 
dwarfs, L2 pec (red), whose red spectrum can not be easily attributed to youth. In addition, we confirm that WISEP J060738.65+242953.4 is 
a very late L dwarf (L$_{\rm o}$8/L$_{\rm n}$9) at the L/T transition using optical and low-resolution NIR spectroscopy. If it remains a 
single object, the presence of ambiguous CH$_{4}$ absorption in the $H$ band represents the earliest onset for any L dwarf at the L/T 
transition in the SpeX Library. Lastly, we provide a transformation from $I_{\rm C}$ to $i$ for L dwarfs using data from \citet{Dahnetal2002} 
and SDSS.

Future work should include parallax measurements to determine distance with more confidence. High resolution imaging/spectroscopy 
is warranted to search for binarity, especially for the very late L dwarfs, WISEP J060738.65+242953.4 and WISE J140533.32+835030.5, where ambiguous CH$_{4}$ 
absorption in the $H$ band of WISEP J060738.65+242953.4 could be the result of an unresolved T dwarf companion. Observations to determine the 
photometric variability and polarization of WISEP J060738.65+242953.4 and WISE J140533.32+835030.5 may reveal signatures of inhomogeneous cloud 
cover \citep{Marleyetal2010}. Close L dwarfs at the L/T transition such as WISEP J060738.65+242953.4 and WISE J140533.32+835030.5 will serve as a proving 
ground to resolve outstanding issues regarding this poorly understood phase of evolution. WISE is beginning to complete the 
census of L dwarfs in the solar neighborhood, most notably, L dwarfs at the L/T transition within 10 pc.

\section{ACKNOWLEDGMENTS}
We thank the anonymous referee for a thorough report that helped to improve the manuscript.
We thank the Annie Jump Cannon Fund at the University of Delaware for support. 
This publication makes use of data products
from the Wide-field Infrared Survey Explorer, which is a joint project of the University of California,
Los Angeles, and the Jet Propulsion Laboratory/California Institute of Technology, funded by the National
Aeronautics and Space Administration.
This publication makes use of data products from
the Two Micron All Sky Survey, which is a joint project of the University of Massachusetts and the
Infrared Processing and Analysis Center/California Institute of Technology, funded by the National
Aeronautics and Space Administration and the National Science Foundation.
Funding for SDSS-III has been
provided by the Alfred P. Sloan Foundation, the Participating Institutions, the National Science Foundation,
and the U.S. Department of Energy. SDSS-III is managed by the Astrophysical Research Consortium for the
Participating Institutions of the SDSS-III Collaboration including the University of Arizona, the
Brazilian Participation Group, Brookhaven National Laboratory, University of Cambridge, University of
Florida, the French Participation Group, the German Participation Group, the Instituto de Astrofisica de
Canarias, the Michigan State/Notre Dame/JINA Participation Group, Johns Hopkins University, Lawrence Berkeley
National Laboratory, Max Planck Institute for Astrophysics, New Mexico State University, New York University,
Ohio State University, Pennsylvania State University, University of Portsmouth, Princeton University, the
Spanish Participation Group, University of Tokyo, University of Utah, Vanderbilt University, University
of Virginia, University of Washington, and Yale University.
Some of the data presented herein were obtained at the W.M. Keck Observatory, which
is operated as a scientific partnership among the California Institute of Technology, the
University of California and the National Aeronautics and Space Administration. The
Observatory was made possible by the generous financial support of the W.M. Keck
Foundation. The authors wish to recognize and acknowledge the very significant cultural
role and reverence that the summit of Mauna Kea has always had within the indigenous
Hawaiian community. We are most fortunate to have the opportunity to conduct
observations from this mountain.
This research has 
benefitted from the SpeX Prism Spectral Libraries, maintained by Adam Burgasser at 
http://pono.ucsd.edu/\textasciitilde{}adam/browndwarfs/spexprism. 
This research has made use of the NASA/IPAC Infrared Science Archive, which is
operated by the Jet Propulsion Laboratory, California Institute of Technology, under contract
with NASA.
This research has benefitted from the M, L, 
and T dwarf compendium housed at DwarfArchives.org and maintained by Chris Gelino, Davy Kirkpatrick, 
and Adam Burgasser. 
This research has made use of the VizieR catalog access tool, CDS, Strasbourg, France. 
This research has made use of the SIMBAD database, operated at CDS, Strasbourg, France. The Digitized 
Sky Surveys were produced at the Space Telescope Science Institute under U.S. Government grant NAG W-2166. 
The images of these surveys are based on photographic data obtained using the Oschin Schmidt Telescope 
on Palomar Mountain and the UK Schmidt Telescope. The plates were processed into the present compressed 
digital form with the permission of these institutions.

\bibliographystyle{apj}
\bibliography{/Users/Phil/research/bibliography}

\label{lastpage}

\end{document}